\documentclass[]{pasj02} 
\usepackage[switch,mathlines]{lineno} 
\usepackage{url}
\usepackage{booktabs}
\usepackage{afterpage} %
\usepackage{lmodern}
\usepackage{fix-cm}
\usepackage[whole]{bxcjkjatype}
\usepackage[normalem]{ulem}
\usepackage{amsmath}
\usepackage{geometry}
\usepackage{enumitem}
\usepackage{float}
\usepackage{caption} 
\usepackage{graphicx}
\usepackage{tikz}
\usetikzlibrary{shapes.geometric, arrows.meta, positioning}

\usepackage{cuted}   
\usepackage{capt-of}  
\usepackage{tabularx}
\usepackage{makecell}
\usepackage{multirow}
\usepackage{placeins}
\usepackage{overpic}
\usepackage{xcolor}


\renewcommand{\textbf}[1]{{#1}}
\newcommand{\jm}[1]{{}}
\newcommand{\jmold}[1]{{}}

\newcommand{\SGR}{Sgr~A$^{*}$}

\newcommand{\DEG}{^{\circ}}

\newcommand{\kuro}{\color{black}}
\newcommand{\aoi}{\color{blue}}

\newcommand{\UVC}{\textit{u-v}~}
\newcommand{\GHZ}{GHz~}
\jyear{2025}
\Received{2025/05/20}
\Accepted{2026/04/20}

\begin{document} 
\title{Short timescale variation in the submillimeter flux of Sagittarius~A$^{\ast}$}
%
\author{
Makoto \textsc{Miyoshi},
\altaffilmark{1}\altemailmark\orcid{0000-0002-6272-507X}
\email{makoto.miyoshi@nao.ac.jp}
Yoshiaki \textsc{Kato},
\altaffilmark{2}\altemailmark\orcid{0000-0003-2349-9003}
\email{yoshi\_kato@met.kishou.go.jp}
Yoshiharu \textsc{Asaki},
\altaffilmark{3}\altemailmark\orcid{0000-0002-0976-4010}
Masato \textsc{Tsuboi},
\altaffilmark{4}\altemailmark\orcid{0000-0001-8185-8954} 
\email{masato.tsuboi@meisei-u.ac.jp} 
Kenta \textsc{Uehara},
\altaffilmark{5}\altemailmark\orcid{0000-0001-9264-5620}
Tomoharu \textsc{Oka},
\altaffilmark{6}\altemailmark\orcid{0000-0002-5566-0634}
Masaaki \textsc{Takahashi},
\altaffilmark{7}
Jos\'e K. \textsc{Ishitsuka},
\altaffilmark{8}\altemailmark\orcid{0000-0003-3940-0127}
Takahiro \textsc{Tsutsumi},
\altaffilmark{9}\altemailmark\orcid{0000-0002-4298-4461}
\email{ttsutsum@nrao.edu} 
Atsushi \textsc{Miyazaki},
\altaffilmark{10}
\email{atsmiyazaki2@a-net.email.ne.jp}
and~
Ryoji \textsc{Matsumoto}
\altaffilmark{11}\altemailmark\orcid{0000-0002-9500-384X}
}
\altaffiltext{1}{National Astronomical Observatory of Japan (NAOJ), Mitaka, Tokyo 181-8588, Japan}
\altaffiltext{2}{Japan Meteorological Agency: 3-6-9 Toranomon, Minato City, Tokyo 105-8431, Japan} 
\altaffiltext{3}{
Joint ALMA Observatory, Alonso de C\'{o}rdova 3107, Vitacura, Santiago, 763 0355, Chile;
National Astronomical Observatory of Japan, Alonso de C\'{o}rdova 3788, Office 61B, Vitacura, Santiago, Chile;
Department of Astronomical Science, School of Physical Sciences, The Graduate University for Advanced Studies (SOKENDAI), 2-21-1 Osawa, Mitaka, Tokyo 181-8588, Japan
}
\altaffiltext{4}{School of Science and Engineering, Meisei University, 2-1-1 Hodokubo, Hino, Tokyo 191-8506, Japan}
\altaffiltext{5}{Department of Astronomy, The University of Tokyo, Bunkyo, Tokyo 113-0033, Japan}
\altaffiltext{6}{
School of Fundamental Science and Technology, Graduate School of Science and Technology, Keio University, 3-14-1 Hiyoshi, Kohoku-ku, Yokohama, Kanagawa 223-8522, Japan
}
\altaffiltext{7}{Department of Physics and Astronomy, Aichi University of Education, Kariya 448-8542, Japan}
\altaffiltext{8}{Faculty of Electric and Electronic Engineering, National University of the
Centre of Peru, Av. Mariscal Castilla No. 3909, El Tambo, Huancayo, Peru.}
\altaffiltext{9}{National Radio Astronomy Observatory, Socorro, NM 87801-0387, USA}
\altaffiltext{10}{Japan Space Forum, Kanda-surugadai, Chiyoda-ku, Tokyo,101-0062, Japan}
\altaffiltext{11}{Department of Physics, Graduate School of Science, Chiba University, 1-33 Yayoi-Cho, Inage-Ku, Chiba 263-8522, Japan}
%

\maketitle
\begin{abstract}
We present a study of short-timescale 340~GHz flux-density variability of Sagittarius~A$^{*}$ (\SGR) using ALMA Cycle~3 observations.
Careful self-calibration enabled snapshot imaging with 10~s integrations, achieving an effective image-domain SNR exceeding $10^{8}$ and allowing high-cadence monitoring of major Galactic Center sources.
To suppress common-mode atmospheric and instrumental fluctuations, we measured the flux of \SGR\  relative to multiple non-variable sources within the same field of view.
We further quantified and corrected apparent variability induced by time-dependent \UVC coverage and the associated PSF changes by performing imaging simulations with a static input model.
These procedures isolate the intrinsic intensity variations of \SGR\ with substantially reduced non-source contamination.
We searched for characteristic variability timescales over $20~\mathrm{s}<\tau\le T_{\mathrm{obs}}/3$ using structure-function analysis, the Lomb--Scargle method, and state-space-model-based autoregressive spectral analysis.
No dominant narrow periodic component is detected. Instead, we identify a distinct short-timescale flat (white-noise--like) regime at $\tau \lesssim 2.3$--$6.3~\mathrm{min}$, followed by red-noise--like behavior at longer timescales. The short-timescale white-noise regime appears in both active and quiescent phases, indicating statistically independent fluctuations on these timescales. 
We interpret the upper boundary of the white-noise regime as an empirical transition timescale below which fluctuations remain effectively decorrelated and above which temporally correlated variability emerges. Existing theoretical and numerical studies of black-hole accretion flows have not, to our knowledge, explicitly predicted a physically white power spectrum at the shortest timescales, although they do not necessarily exclude such behavior. 
Reported results have more commonly shown red-noise--like or broken-power-law variability, and the physical origin of the flat short-timescale component identified here therefore remains uncertain.
\end{abstract}
\section{Introduction}\label{sect:intro}
~Sagittarius A$^{*}$~(\SGR) is a compact radio source located at the dynamical center of the Galaxy and is associated with a supermassive black hole of mass $M_{\rm BH} \sim 4 \times 10^6~M_{\odot}$~\citep{Ghez:08, Gill:09, Schodel:09, Boehle:2016}. 
Owing to its proximity and mass, \SGR\ is regarded as one of the best laboratories for studying black hole physics and relativistic effects in strong gravitational fields.\\
~At X-ray and near-infrared wavelengths, \SGR\ shows prominent flare-like variability on timescales of approximately 30 min, often accompanied by large amplitude changes~\citep{baganoff2001, Genzel:03, ghez2004}. 

Similar variability has also been observed at millimeter and submillimeter wavelengths.
\citet{Miyazaki:04} first reported intra-day variability at 100 and 140~GHz.
Subsequent long-term monitoring revealed a characteristic mean flux density of $\sim$3~Jy
with variations of 30--50\% on hourly timescales~
\citep{zhao2003, eckart2008sim, marrone2008},
 and these results were further investigated in later studies~\citep{Dexter:14,Witzel+18}.
In the past decade, the emergence of ALMA, with its unprecedented sensitivity, has opened a new phase in studies of flux density variability of Sagittarius~A$^{*}$ at millimeter and submillimeter wavelengths~\citep{Iwata2020,Witzel+21,MW2021,Wielgus:22}.
These observations have established the presence of significant intensity variations on timescales shorter than about an hour.

While short-timescale flux density variability of Sagittarius~A$^{*}$ at millimeter and submillimeter wavelengths has now been firmly established, its accurate measurement still leaves room for improvement.
For example, \citet{Wielgus:22} took advantage of ALMA's high sensitivity to observe \SGR\ at 230~GHz, confirming significant intensity variations on timescales shorter than an hour.Combined with simultaneous observations with the SMA, \citet{Wielgus:22} demonstrated that the previously reported variations indeed originate from \SGR.
However, a discrepancy of up to 0.4~Jy between the flux densities measured by the two arrays remains, suggesting that uncertainties in absolute intensity measurements persist in the millimeter and submillimeter bands.

~\citet{Tsuboi:16} presented images of the Galactic Center region with an extremely high dynamic range  over $2\times10^4$ in the resultant maps of the 250 and 340 GHz bands  using ALMA Cycle~0 data. In the study, self-calibration techniques and/or hybrid mapping methods~\citep{RefHM3,Cornwell:1981} developed for VLBI imaging were applied to ALMA data. 

Using the same technique, we found that it was possible to detect not only \SGR\ but also surrounding objects within the same field of view from data with only 10 seconds of integration in ALMA Cycle~3.
~The snapshot maps not only allow us to follow the ~intensity variations of \SGR\ with high time resolution.
By referencing the flux densities of surrounding objects in the same field of view~(FOV), we can obtain a relative intensity measurement of \SGR, i.e. their ratios. Since the atmospheric and instrumental components in the intensity variations are common to both, they cancel each other out, allowing us to obtain intensity variations due solely to \SGR\ itself. Reporting the results is one objective of this paper.

~Another objective of this paper is to report the results of applying state-space modeling to the time series of \SGR\ intensity variations. 

Not limited to \SGR\ observations, astronomical observation data is often irregularly sampled.  Frequently, applying a simple Fourier transform to it yields erroneous spectra.
 The Lomb-Scargle method (LSM)~\citep{Lomb:76,Scargle:82}
has long been a standard tool for analyzing unevenly sampled time series data, but it suffers from several critical problems.
It cannot be applied to constant-value time series because it depends on the data variance~\citep{VanderPlas:18}, and it is still highly sensitive to irregular sampling, which can produce artificial peaks in the periodogram~\citep{Scargle:82}. 
Furthermore, LSM assumes white Gaussian noise, whereas real astronomical data often exhibit red-noise or systematic trends \citep{Reegen:07}, which can bias the detection of periodic signals. 
The associated False Alarm Probability (FAP), which gives us a measure of the reliability of LSM results, is also prone to inaccuracies \citep{Horne:86}. 
In addition, standard LSM assumes a zero-mean, which can lead to bias if the data contains trends or non-zero mean values, although this problem is partially addressed by Generalized LSM (GLS) \citep{Zechmeister:09}.
We then considered a state-space modeling (SSM)  approach as a more robust alternative under irregular sampling and observational noise.
In particular, the SSM framework enables us to characterize intrinsic variability and to assess candidate characteristic timescales without relying on strict periodicity.\\

~In this paper, we present a time-series analysis of short-timescale flux variations in \SGR\ at 340 GHz using 10.08-second snapshot imaging. We examine both the statistical properties and  characteristic variability timescale  of the variability through an approach that combines differential photometry and state-space modeling.
The structure of this paper is as follows:
Section~\ref{sect:Obs} details the observational data and calibration procedures, along with the method for producing high-time-resolution snapshot maps.
In Section~\ref{sect:Res}, we present the results of our analysis. These results include light curves, structure function analysis, and 
 characteristic variability timescale  search using a state-space model.
In Section~\ref{sect:DISC}, we  discuss whether the short-period white noise-like fluctuations detected in the analysis originate from intrinsic \SGR characteristics or observational errors. If the feature is intrinsic to \SGR, we estimete its physical significance. 
~Finally, Section~\ref{sect:CONC} summarizes the main findings and discusses future prospects.

In addition, Appendix~\ref{Sec:CLEAN-sim} presents imaging simulations using a static input model to quantify and correct apparent snapshot-to-snapshot flux variations caused by time-dependent $uv$ coverage and PSF changes.
Appendix~\ref{Sec:unevensample} investigates how irregular sampling intervals affect the spectral analysis, thereby validating the robustness of the inferred variability properties.

\section{Data Analysis} \label{sect:Obs}
In this section, we describe the observational data and our procedures for data calibration and imaging. Section~\ref{Sec:datacal} outlines the calibration process, and Section~\ref{Sec:calmap} describes the creation of snapshot maps. 
In Section~\ref{Sec:rawJy}, we explain how we extracted the flux densities from the images to quantify the intensity variations of \SGR\ relative to nearby sources - a quantity hereafter referred to as the \textit{relative flux density}.
\subsection{Data calibration}\label{Sec:datacal}
 We analyzed the 340~\GHZ continuum data observed as an ALMA~Cycle~3 program (2015.1.01080.S.; P.I.: M. Tsuboi, \cite{Tsuboi:17}). 
 The observations were performed on four separate days, namely April 23, August 30/31, and September 8 in 2016, spanning 138~days.

 After the standard data calibration processing prescribed by ALMA, we further performed the same calibrations that \cite{Tsuboi:16} performed using the NRAO Astronomical Image Processing System (AIPS).
First, we removed the residual fringe rates and delays of the visibility by using the task FRING, which was found to be quite effective for removing the rapid phase fluctuations caused by atmospheric variations in the given data. 

Second, we attempted to minimize the residual complex gain errors of the data using the self-calibration method~\citep{RefHM3,Cornwell:1981} by the task CALIB and performed the Hybrid mapping process.
In the standard hybrid mapping process, a single point source is used as the initial image model for self-calibration and the refinement of the image model is continued with replacing trial imaging iteratively.
However, we found that the usage of the Galactic Center map obtained previously~\citep{Tsuboi:16} as the image model provided most reliable solutions of calibration.
Given the observed visibility data, which indicated a mean flux density of approximately 4~Jy, we developed a model image by adjusting the flux density at the ~position in the 340~GHz image of~\citet{Tsuboi:16} using the task  \texttt{IMMOD}  in AIPS. 
This image model was then utilized in the self-calibration process. We then obtained the calibration solutions of both amplitude and phase to get the final calibrated visibility data.

Data calibration using amplitude solutions derived from self-calibration was essential to recover the fine structure of the Galactic Center. 
Although, in general, excessive use of amplitude self-calibration should be avoided, we found that applying only the phase solutions was insufficient to recover the fine structure of the Galactic Center from the data.
As a test, we applied self-calibration solutions derived from the image model corresponding to the CLEAN components shown in Figure~\ref{fig:fig01}. Even in this case, where the image model should be highly reliable, the use of phase solutions alone was insufficient to reproduce the image satisfactorily.

\subsection{Making of snapshot maps}\label{Sec:calmap}
\begin{figure*}
\includegraphics[trim={1.25cm 0.25cm 0.25cm 0.25cm},clip,width=\textwidth]{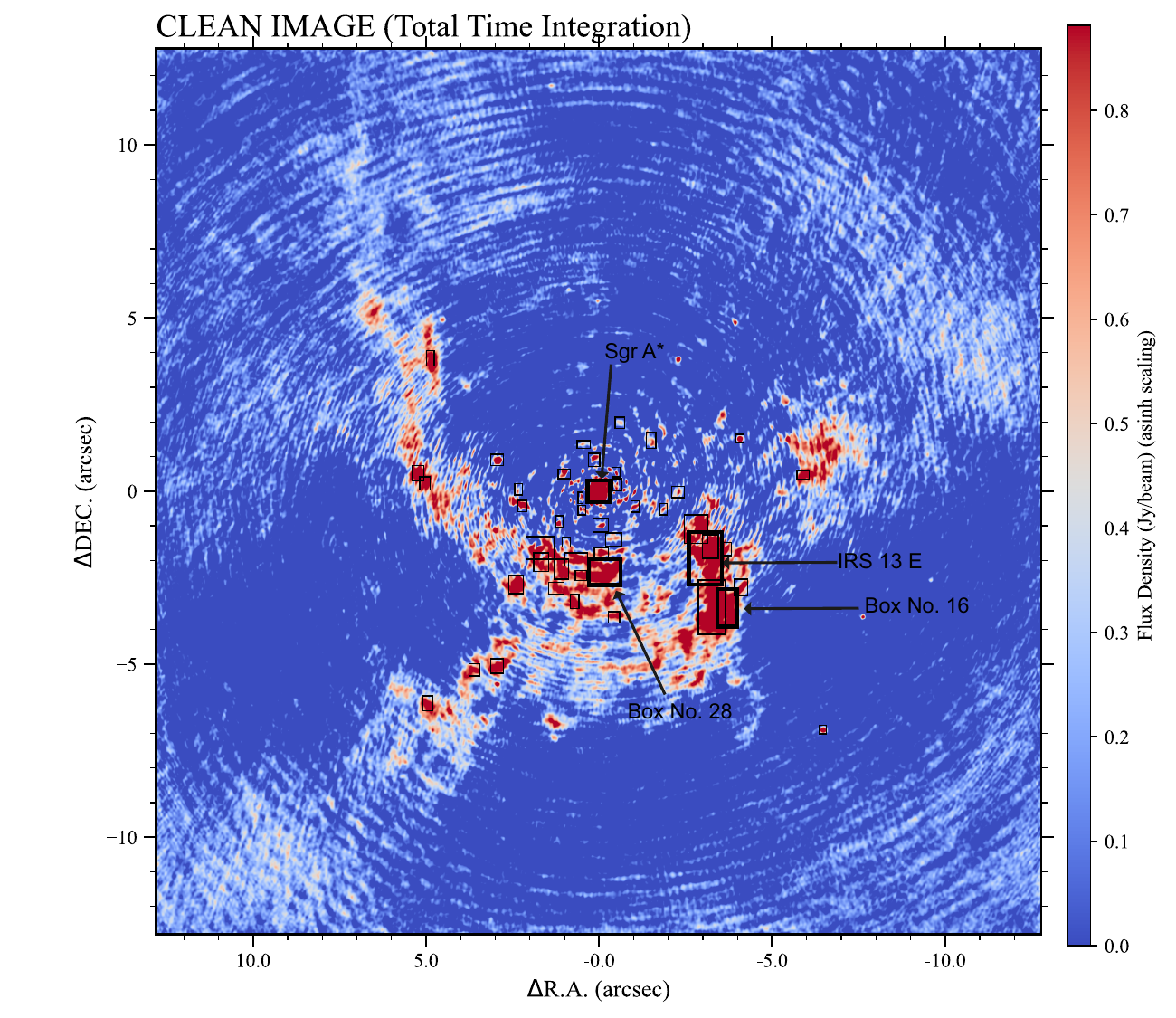}
\caption{ 
 Synthesis image of the Galactic Center region at 340~\GHZ (continuum) using ALMA. 
 All data from the four observing epochs were used. 
The imaging area is a $25"\times 25$" field centered on \SGR.
The unit of the color bar scale  unit  is Jy/beam. 
The FWHM beam size is $0.''24 \times 0.''22$, and $PA = -89.\DEG1$. 
The boxes indicate the  intensity reference source areas  we used.
Using the task IMSTAT in AIPS, we obtained the flux densities in the areas and summed them up to compare with that of \SGR.
Also the measured flux densities of the boxes with labels are shown in Figure~\ref{fig:fig04}.\\{Alt text: ALMA 340~GHz continuum image of the Galactic Center region, produced by combining data from all four observing epochs. The image shows \SGR\ and nearby compact sources such as IRS 13 with the Galactic Center mini-spiral. Small boxes indicate the regions used as reference sources for the relative flux density.}
}
\label{fig:fig01}
\end{figure*}
\begin{figure*}[tb1]
\centering
\includegraphics[trim={0.30cm 2.25cm 0.25cm 2.75cm},clip,width=\textwidth]{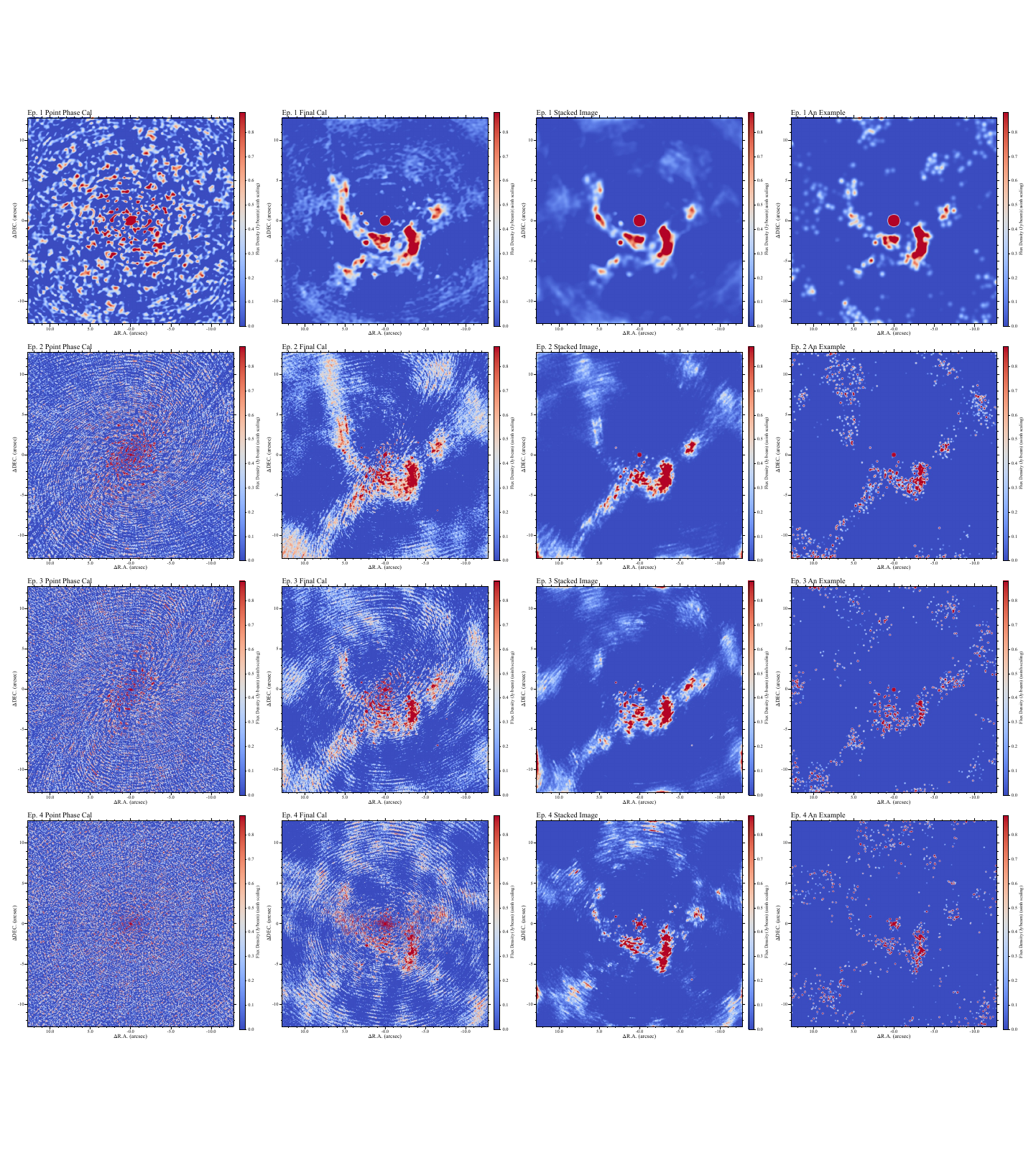}
\caption{
Images from each stage: From left to right:
(1) Images after applying the phase self-calibration solution using the point source model,
(2) Images after applying the final phase and amplitude self-calibration solution, 
(3) Stacked images of CC2IM snapshot maps from each epoch duration, and 
(4) Examples of 10.08-second integration CC2IM snapshot maps (composed only of CLEAN components) are shown.
From top to bottom, images from Epochs 1, 2, 3, and 4 are shown.
}
\label{fig:fig02}
\caption*{\footnotesize\textit{Alt text:} 
The comparison of snapshot images from each observing epoch, showing the effects of different calibration and imaging methods on image quality. Panels show: (1) phase self-calibration using a point source model, detecting only \SGR; (2) hybrid-mapping using refined phase solutions, revealing nearby objects and the mini-spiral structure, with concentric ripple-like artifacts caused by the time variability of \SGR; (3) stacked 10.08-second CC2IM snapshot maps, where ripple artifacts are suppressed; (4) an example of a 10.08-second CC2IM snapshot map, showing nearby sources including the mini-spiral.}
\end{figure*}

~The imaging results obtained by integrating the entire dataset after applying the final self-calibration solutions are shown in Figure~\ref{fig:fig01}. The Galactic Center mini-spiral extending in three directions from the center is clearly visible. Compact radio sources at the Galactic Center, such as IRS~13 and other nearby objects, are also detected.
~Compared to the results of \citet{Tsuboi:17}, we successfully obtained images around~\SGR\ with a higher dynamic range. 
However, a ripple-like structure spreading in concentric circles from the center of \SGR\ is found. 
 Presumably, this feature is an artifact caused by intensity variations in \SGR\ ~during the observation  time. \\
~In addition to the integrated image, we also produced snapshot maps of \SGR\ and the surrounding objects using 10.08-second integrations
(see  the right   panels in Figure~\ref{fig:fig02}).
The ALMA Cycle~3 observations employed 35 antennas, resulting in 595 baselines. Although this is significantly fewer than the 66 antennas and 2145 baselines available in full ALMA operations, the \UVC coverage was sufficient to achieve an adequate dynamic range in the resulting snapshot maps.\footnote{
For reference, the 12-meter array alone consists of 54 antennas, yielding 1431 baselines among them.}
(In contrast, our attempts to produce high-quality snapshot maps using ALMA Cycle~0 data were unsuccessful.)\\

To ensure the reliability of the snapshot flux measurements, the method for producing the maps is described in detail below.
\begin{enumerate}[label=（\arabic*）]
\item
As the observation time progresses, the \UVC coverage of each snapshot gradually changes, leading to variations in the synthesized beam shape and, consequently, the spatial resolution of the snapshot maps. We investigated the extent of these beam fluctuations. 
The default beam shapes defined by the \UVC coverage
\footnote{ In radio interferometry, the “beam” (spatial resolution) is defined by the shape of the main lobe of the dirty beam (PSF); information about the positions and amplitudes of sidelobes in the dirty-beam is not included in that definition.}
for each snapshot are summarized in Table~\ref{tab:beam}.
The fluctuations in beam size were not as large as initially feared. 
The beam width changed by approximately 10~milli-arcseconds (mas) during each observation epoch. 
(Note that Epoch~1 exhibited a particularly large beam size due to the compact configuration of the ALMA array.)
To mitigate the adverse effects of beam shape fluctuations on the flux-density measurements, we adopted a uniform circular beam with a size close to the average beam for each epoch when producing the snapshot maps (see Table~\ref{tab:beam}).
 Note  that comparisons of measured flux densities between different epochs are not straightforward, and even comparisons of intensity ratios across epochs require caution.
Accordingly, we analyze each epoch independently and avoid interpreting 
 
differences in the absolute level of the relative flux density across epochs. 

\item
In the practical deconvolution of a "dirty beam," or the point spread function (PSF), the side-lobe structures in dirty beam cannot be completely removed. 
Artifact structures induced by side-lobes may be incorporated into the deconvolved image. 
Particular caution is required when the target is not a single point source, but rather an extended, structurally complex object. 
The side-lobe levels increase because the \UVC ~coverage of snapshot observations is sparser than that of a full-track observation. Coupled with reduced sensitivity, this can generate spurious temporal variations in the measured intensity. 
We quantified this effect via simulations using a static model of the Galactic Center structure (see Appendix~\ref{Sec:CLEAN-sim}). The effect can produce artificial variability of up to $5\%$. 
Taking this result into account, we corrected the observed time variability for this effect.
Concretely, we corrected the measured relative light curves by normalizing them with the time-dependent response obtained from the static-data simulations. 
Let $S_{\mathrm{sim}}(t)$ be the simulated flux density time series for a \emph{static} source processed with the identical \UVC coverage and imaging pipeline as the real data.
The corrected light curve is then
\begin{equation}
R_{\mathrm{corr}}(t)\;=\;\frac{R_{\mathrm{obs}}(t)}{S_{\mathrm{sim}}(t)}\,.
\end{equation}
where the division is performed snapshot by snapshot.

\item
Snapshot maps have only very short integration times,
 which makes the effects of noise 
more pronounced compared to full-time integrated images.
In standard CLEAN processing, the final image, called the "CLEAN map", is obtained by convolving the restoring beam 
—the result of a Gaussian fit to the main lobe of a dirty beam—with all CLEAN components.
Then, the residual map remaining after CLEAN iterations is added to maintain consistency of the noise level with that of the dirty map.

 Therefore, images composed solely of CLEAN components (hereafter referred to as CC2IM images) are expected to reduce the impact of noise on our measurements.
At the same time, the CLEAN algorithm may leave residual maps that retain a fraction of genuine source structure, and this limitation should be kept in mind. 
To assess this, simulation results using a static image model were employed. 
The CC2IM snapshot images showed approximately $25\%$ lower levels of apparent intensity variation due to \UVC fluctuations compared to the case of CLEAN images.
On this basis, CC2IM snapshot images were adopted for our measurements, as they are less susceptible to the effects of noise than a standard CLEAN map.
\end{enumerate}

To evaluate the quality of the images produced by the above methods, we compared them with trial images generated at  different  stages of processing.
Figure~\ref{fig:fig02} presents the images at each stage. From left to right, the following images are shown:
\begin{enumerate}[label=（\arabic*）]
\item 
Images obtained by 
applying phase-only self-calibration with a point-source model and subsequent CLEAN deconvolution.

\item 
Images obtained by hybrid mapping, 
in which the model image is iteratively refined for self-calibration. The final phase and amplitude self-calibration solutions were applied to the data, followed by CLEAN.

\item 
Stacked images produced by combining all CC2IM snapshot maps in each epoch.

\item
Examples of 10.08-second CC2IM snapshot maps composed 
solely of CLEAN components. For each epoch, the snapshot map corresponding to the central time is shown.
\end{enumerate}
In Figure~\ref{fig:fig02}, the images for Epochs~1, 2, 3, and 4 are presented from top to bottom.
In case (1), only the central source, \SGR, is detected, while the mini-spiral and other surrounding structures in the Galactic Center region remain undetected.
In case (2), the mini-spiral and other structures become detectable. 
However, we also observe concentric ripple-like patterns, which are likely artifacts caused by intensity variations in \SGR. 
In case (3), stacking the 10.08-second integrated images reduces the ripple structures, making them less prominent.
In case (4), since the intensity of \SGR\ remains constant within the 10.08-second integration, ripple structures are not observed. 

\begin{figure}[t]
\centering
\includegraphics[clip, width=\linewidth]{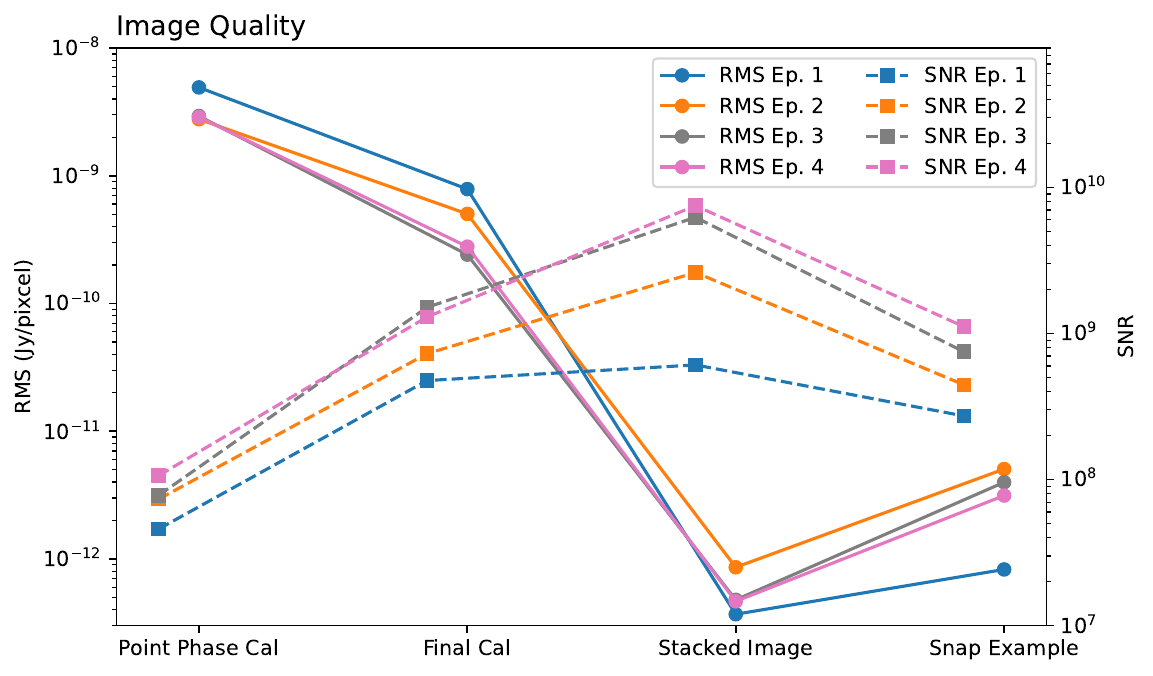}
\captionsetup{width=\columnwidth}
\caption{
Image quality. RMS noise levels and SNRs of the images in Figure~\ref{fig:fig02} are shown. 
The solid line represents the RMS noise (the left Y-axis scale), while the dotted line represents the SNR (the right Y-axis scale).
 Here, the signal-to-noise ratio (SNR) is defined as the ratio of the image peak brightness to the RMS noise level.
RMS noise was measured using the task \texttt{IMSTAT} in AIPS within the square of $25.56~\rm arcsec~(R.A.~direction) \times 3.44~\rm asec~(\delta~direction)$ at the southern edge of the image.
\\ {Alt text: Comparison of image RMS noise and signal-to-noise ratio (SNR) for the image types shown in Figure~\ref{fig:fig02}. Left axis: RMS noise in Jy per pixel. Right axis: SNR. The results demonstrate the significant noise reduction achieved through advanced calibration and CLEAN-component-only imaging.}
}
\label{fig:fig03}
\end{figure}

Additionally, CC2IM images, produced \ solely from CLEAN components, can reduce RMS image noise.
A quantitative evaluation of the image quality is presented in Figure~\ref{fig:fig03}.
The RMS noise level in case~(3) is reduced by nearly four orders of magnitude relative 
to that in case~(1).
The RMS noise level of the 10.08-second snapshot maps in case~(4) is about one order of magnitude higher than that of the stacked images in case~(3), which is expected 
given the shorter integration time.
In this comparison, the CC2IM  snapshot maps used for the relative flux-density measurements appeared to have an RMS noise level approximately two orders of magnitude lower than in case~(2), with an integration time shorter by roughly two orders of magnitude.
\subsection{Flux density from the obtained image of \SGR}\label{Sec:rawJy}
~Radio continuum emissions at 340~GHz around \SGR\ are predominantly from interstellar and/or circumstellar dust, and partially from ionized gas in H II regions~\citep{Tsuboi:16, Paumard2006, zhao2009, Genzel1996}.
These sources do not show intrinsic intensity variations on timescales of several hours. By referencing their stable flux densities, we can achieve reliable measurements of the intrinsic intensity variations of \SGR.
The  ideal  reference is an isolated, non-variable point source within FOV. 
However, because Galactic Center images contain numerous compact and extended objects, such an isolated point source is not available. 
We therefore adopt as the reference the sum of the flux densities measured over selected compact BOX regions.
A summary of the simulations performed to examine this issue is provided in Appendix~\ref{Sec:CLEAN-sim}.

To ensure the reliability of our measurements, we next describe two key 
 properties of the intensity measurements. Understanding these properties is essential; neglecting them may lead to misinterpretation of the results.

 First, the measured values depend on the spatial resolution in each observing epoch.
The measured flux density of \SGR\ at 340~\GHZ was approximately 4~Jy, whereas 
that of the IRS~13 complex—the second-brightest object in the field—was 
$30$–$40$~mJy.
The integrated flux density of the imaging area excluding \SGR\ was about $1.2$~Jy~(Epoch~1) or $0.13$~Jy~(Epochs~2-4). 
Differences across epochs arise because most sources in the field are extended, and the measured flux density depends on the interferometer's spatial resolution at each epoch.
(Note that \SGR\ can be regarded as a point source when compared with the ALMA's spatial resolution, so the measured flux densities of \SGR\ 
are largely insensitive to the spatial resolution.)
 Therefore, the (relative) flux-density values of \SGR\  should not be directly compared across epochs without careful consideration.

The second point concerns the use of self-calibration. 
Self-calibration is a long-standing technique for calibrating VLBI data. When applied appropriately, it stabilizes both amplitudes and phases, yielding high-quality images. 
Recently, it has also been successfully adopted for data calibration in connected-element radio interferometers, such as ALMA.

\begin{figure*}[t]
\centering
\includegraphics[trim={0cm 0cm 0cm 0cm},clip,width=\textwidth,height=170mm]{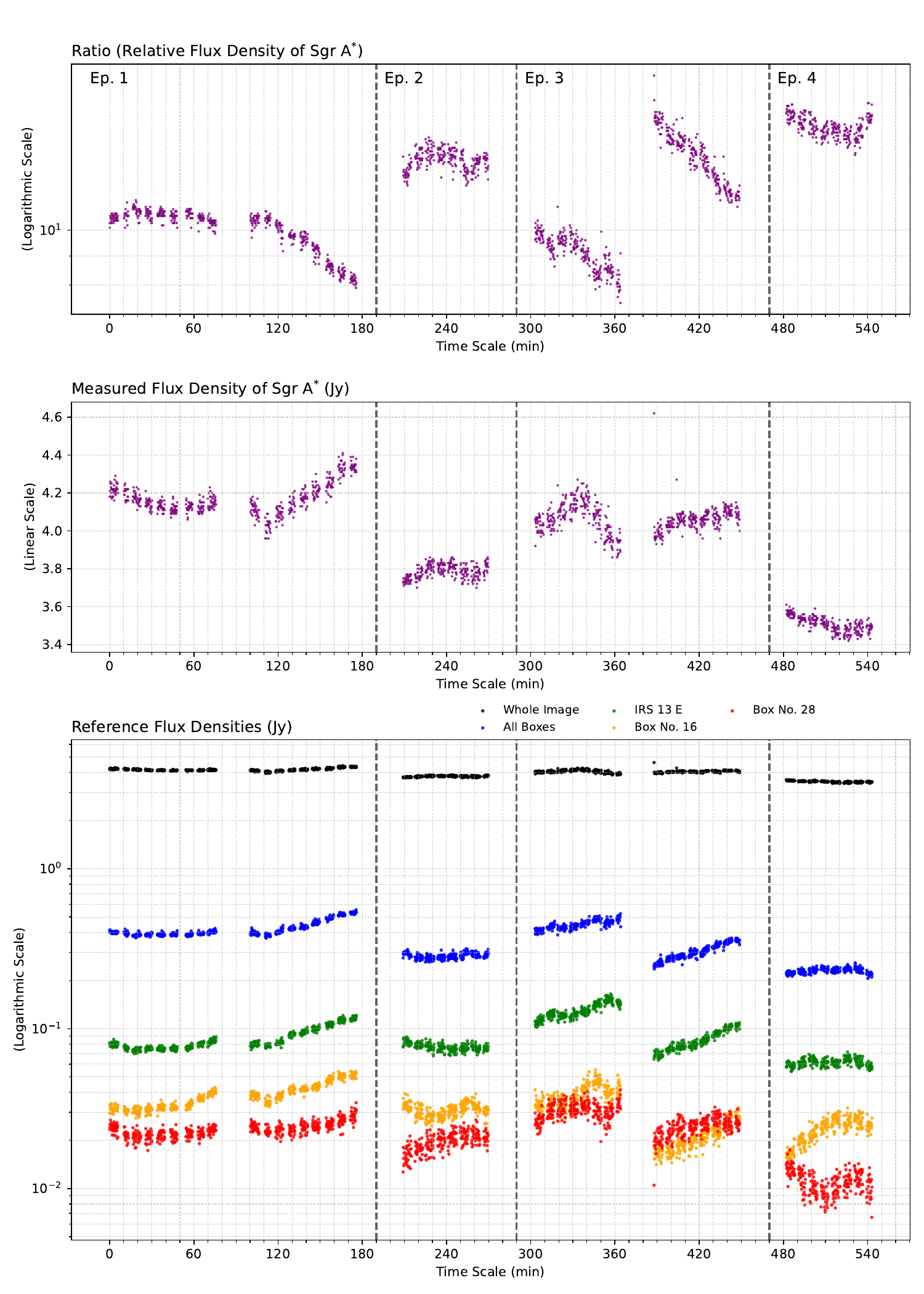}
\caption{
Measurements from CC2IM snapshots.
Measurements from snapshot CC2IM images reconstructed from observed visibility data. 
The CC2IM images consist solely of CLEAN components. 
Top: Relative flux density of \SGR, defined as the ratio of the flux density of \SGR\ to that of the reference sources. 
Middle: Measured flux density of \SGR. 
Bottom: (1) the sum of the measured flux densities of the reference box areas, 
(2) those of selected box areas (see Fig.~\ref{fig:fig01}), and 
(3) the measured flux density of the entire map. 
Note that no correction for apparent variability induced by \UVC\ coverage has been applied in this plot (raw light curves). 
All subsequent analyses (SF and spectral analyses) use the \emph{bias-corrected} relative light curves described in Section~\ref{Sec:calmap}.
\textit{Alt text:} Measured flux densities from snapshot CC2IM images reconstructed from observed visibility data. 
The CC2IM images consist solely of CLEAN components. 
Top: relative flux density of \SGR\, showing the variability. 
Middle: measured flux density of \SGR\, which remains nearly constant. 
Bottom: flux densities of selected reference sources and total image flux, showing pronounced variability. Note that no correction has been applied for fluctuation arising from changes in u–v coverage in this plot.
}
\label{fig:fig04}
\end{figure*}

However, researchers using self-calibration should recognize that applying self-calibration solutions leads to certain losses of information in the resulting images.
\begin{enumerate}[label=（\arabic*）]
\item
It is well known that applying phase self-calibration removes the absolute astrometric information of the target. 
This occurs
because the centroid of the model brightness distribution defines the image origin (phase center). In this work we do not use positional information, so the loss of absolute position is not critical.
\item
 A less widely recognized effect arises from applying amplitude solutions. 
The flux density in the resulting image can be strongly influenced by the model flux density used during self-calibration, biasing it away from the source's true value. 
In other words, the image loses information on the absolute flux scale after applying amplitude self-calibration solutions.
 
 As shown in Figure~\ref{fig:fig04}, the effect on flux density described above can be clearly seen in our snapshot maps when amplitude self-calibration is applied. While the measured flux density of the entire image and that of \SGR, which dominates the field, remain nearly constant, large variations are observed in the measured flux densities of the reference sources, which are expected to be stable.

Amplitude self-calibration removes the absolute flux scale in the resulting images, so analyses relying solely on absolute flux densities are not appropriate.
However, to first order the simultaneously measured in-field intensity ratios are often well preserved within the small FOV.
 Accordingly, we base our analysis on relative (ratio) measurements. 
In practice, we use the intensity ratio between \SGR\ ~and nearby compact references within the same FOV, which suppresses slow, common-mode atmospheric/instrumental drifts and makes the loss of absolute flux density information immaterial for our purposes.
\end{enumerate}
We produced 1620 snapshot maps and summed up the flux densities of the surrounding objects in the 46~boxes shown in Figure~\ref{fig:fig01}. 
The box areas cover, for example, regions of IRS~13, IRS~34 SW, IRS~2, AF, IRS~9W, IRS~21, Her~N2, IRS~1W, and IRS~16NE. Using the task IMSTAT in AIPS, we obtained the flux densities in the box areas and summed them up. Using the method explained above, we thus obtain the intensity variation of \SGR\ largely free from instrumental and atmospheric amplitude fluctuations. 
%
\begin{table*}[t]
\caption{Beam Shapes of Snapshot Maps}
\label{tab:beam}
\begin{tabular}{lccc}
\toprule
Epoch& Major$\times$Minor~(HPBW) &PA ($^\circ$)&Adopted Circular Beam~(HPBW) \\ 
\midrule
Ep.~1& $(0."5053\pm0."0136)\times(0."4539\pm0."0113)$&$-81.1 \pm12.5$&$0."48$\\
Ep.~2& $(0."1749\pm0."0075)\times(0."1207\pm0."0008)$&$-67.3 \pm~1.5$&$0."15$\\
Ep.~3& $(0."1424\pm0."0112)\times(0."1186\pm0."0009)$&$-75.8 \pm~3.7$&$0."13$\\
Ep.~4& $(0."1220\pm0."0005)\times(0."1141\pm0."0005)$&$~~18.6\pm~6.9$&$0."12$\\
\bottomrule
\end{tabular}
\end{table*}

\clearpage
\FloatBarrier
\section{Results}\label{sect:Res}
This section presents the primary findings from our analysis.
First, we present the light curves~(Section~\ref{sect:lcurve}), followed by a statistical characterization of the variability using flux distributions and structure functions~(Section~\ref{sect:hist}). 
We then examine 
 the possible presence of  characteristic variability timescale  
 using both traditional and state-space–modeling (SSM) methods (Section~\ref{sect:qpo}).
For clarity, we reiterate that the intensity of \SGR\ is expressed relative to the summed intensities of nearby sources. This measure is called the relative flux density.
%
\subsection{Light Curves}\label{sect:lcurve}
\begin{figure}[H]
\centering
\includegraphics[trim={3mm 0mm 1mm 0mm}, clip, width=\if@twocolumn\linewidth\else\textwidth\fi]{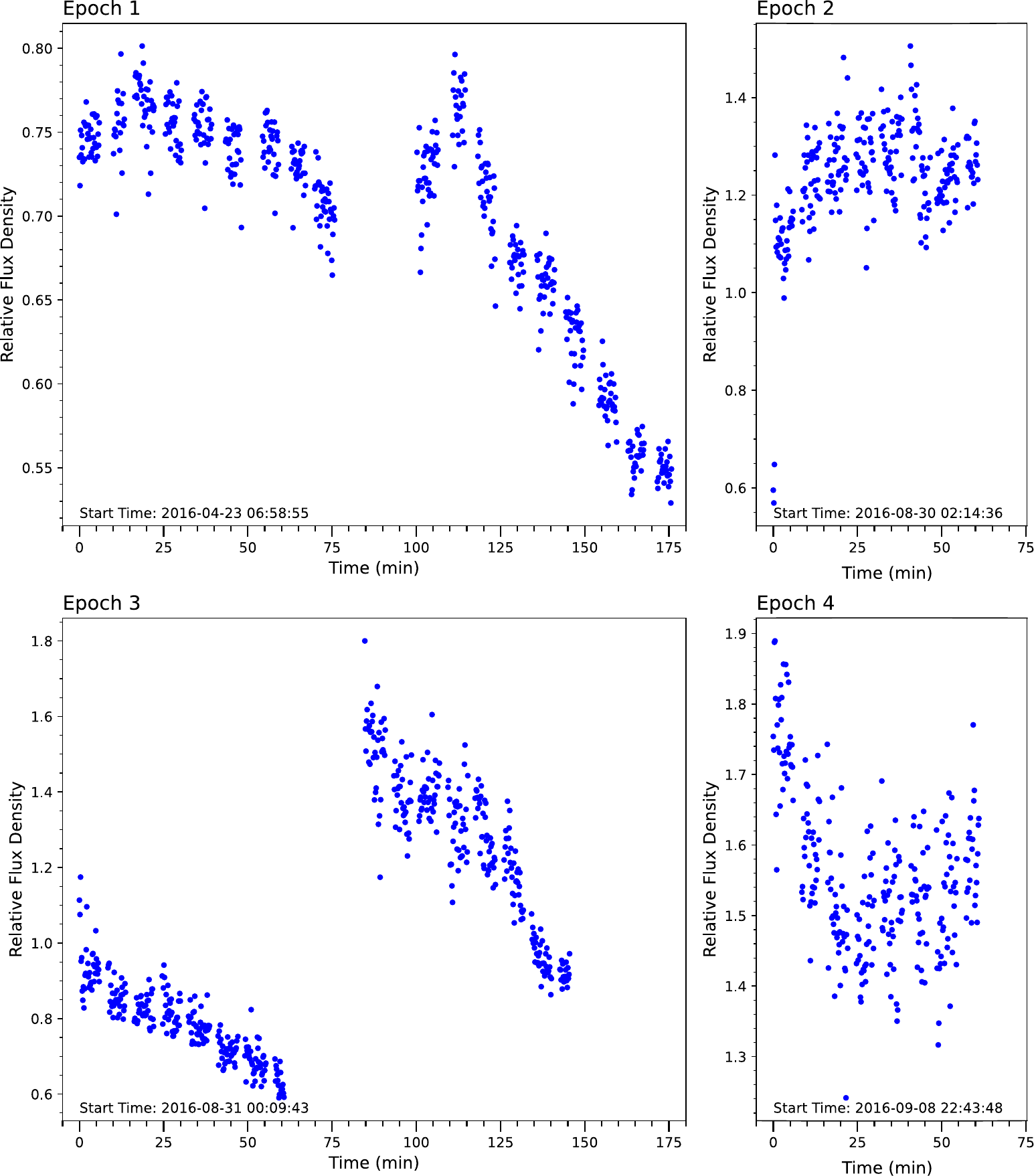}
\caption{
Relative flux density variations of \SGR\ during the four observing epochs.
The subpanels show the variations for Ep.~1, Ep.~2, Ep.~3, and Ep.~4 in order.
The horizontal axis represents time in minutes, and the vertical axis represents the relative flux density of \SGR\ compared to surrounding reference objects. 
Note that although all ratios are referenced to the total flux densities within the same BOX areas, differences in restoring beams between epochs necessitate caution when comparing values across epochs.
Additionally, the origin of the vertical axis is not set at zero.
\\ {Alt text: Time series plots of the measured flux densities for \SGR\ and reference regions. Top: relative flux density of \SGR, showing the variability. Middle: measured flux density of \SGR, which remains nearly constant. Bottom: Flux densities of selected reference sources and total image flux, showing pronounced variability.}
}
\label{fig:fig05}
\end{figure}

The light curves of \SGR~are shown in Figure~\ref{fig:fig05}.
The variability has been corrected for apparent intensity fluctuations caused by \UVC coverage variations.
In the uncorrected (“raw”) light curve shown in Figure~\ref{fig:fig04}, the observed amplitudes exceed those expected from the static-model \UVC\ effect by factors of several up to nearly ten; for the static-model results  (as shown Figure~\ref{fig:fig10}, in Appendix~\ref{Sec:CLEAN-sim}).
These results indicate that the 340~\GHZ variability of \SGR\ is intrinsic; variations on timescales shorter than one hour are evident, with distinct patterns in each epoch.

Although no flare-like activities, as observed in X-ray or near-infrared emissions, were found during our observing  time, the relative flux density of \SGR\ exhibited significant variability.

In Ep.~1, a concave-down trend \ is seen from the start to about $75~\mathrm{min}$; a second rise occurs near $100~\mathrm{min}$, peaking at $110~\mathrm{min}$, followed by a somewhat rapid \ decline toward the end of the observation.
In Ep.~2, the relative flux density increases rapidly at first and then decreases gradually over time.
In Ep.~3, the relative flux density decreases gradually at first; a rapid rise is inferred during the observation gap from $65$ to $80~\mathrm{min}$. 
 After the gap, the flux density decreases again.
In Ep.~4, the relative flux density first decreases, then remains nearly constant from about $25$ to $50~\mathrm{min}$.
~The general characteristics of the light curves are summarized in Table~\ref{tab:tb01}, which lists the observing parameters for each epoch together with a brief description of the relative flux-density variations.
They indicate that the detected variations do not grow linearly with observing time, but instead exhibit variations on short timescales of less than an hour.

The magnitude of the observed variability is consistent with previous measurements at submillimeter wavelengths \citep{zhao2003, eckart2008sim, marrone2008}.
Although earlier  studies inevitably faced challenges from atmospheric and instrumental systematics in absolute flux-density calibration, their reported variability characteristics are broadly consistent with our results, suggesting that they successfully captured intrinsic variability of \SGR.
By measuring \SGR\ relative to multiple nearby, non-variable sources within the same FOV, our approach suppresses common-mode fluctuations and provides a more robust estimate of the intrinsic flux-density variations. The agreement between the independent methods therefore strengthens the conclusion that the measured variability is representative of \SGR\ itself.
\makeatletter
\makeatother
\begin{table*}
\begin{flushleft}
\centering
\begin{tabular}{lcccc} 
\toprule
Observation Date$^{1}$ & Duration & Relative Intensity$^{2}$ & Max/Min$^{3}$ & Volatility\\ 
& \small{(hour)} & & & \small{(hour$^{-1}$)} \\ 
\midrule
Ep.~1:~$2016~Apr.~23$~(~~0~days)&2.93 & $0.70 \pm0.07$& $0.80/0.53~(151.5\%)$& $0.52$\\
Ep.~2:~$2016~Aug.~30$~(129~days)&1.01 & $1.24 \pm0.11$& $1.51/0.57~(264.7\%)$& $2.62$\\
Ep.~3:~$2016~Aug.~31$~(130~days)&2.43 & $1.03 \pm0.29$& $1.80/0.59~(305.1\%)$& $1.26$\\
Ep.~4:~$2016~Sep.~08$~(138~days)&1.02 & $1.56 \pm0.11$& $1.89/1.24~(152.2\%)$& $1.49$\\
\bottomrule
\end{tabular}
\caption{
Observational epochs and relative flux density variations of \SGR. $^{1}$
Observing date; the number in parentheses indicates the elapsed days since the first observation. $^{2}$
Mean relative flux density and standard deviation within each epoch. 
Note that data for Epoch~1 were obtained with a compact array (lower spatial resolution), while data for other epochs were obtained with a larger array (higher spatial resolution). 
Due to the differences in spatial resolution and reference source fluxes, direct comparisons of the relative flux densities between epochs are not meaningful. $^{3}$
Ratio of the maximum to minimum relative flux density within each epoch. $^{4}$
The Max/Min ratio divided by the observing duration (hr), indicating the average flux change rate~(h$^{-1}$).
}
\label{tab:tb01}
\end{flushleft}
\end{table*}
\subsection{ Search range of variability timescales }
\label{sect:sarea}
In the following sections, we investigate variability in the flux density of \SGR\ using structure-function analysis, LSM, and SSM.
Before presenting these results, we first define the range of timescales over which such analyses can be regarded as statistically meaningful.

Because any observational dataset has a finite duration $T_{obs}$, the range of variability timescales that can be reliably probed is inevitably limited. 
This constraint is intrinsic to finite-length time series and does not depend on the specific analysis method employed.

From the perspective of statistical signal processing, the achievable frequency resolution for a time series of length $T_{obs}$ is fundamentally limited to $\Delta f \sim 1/T_{obs}$ \citep{Kay1988,Kay1993,Priestley1981}.
Consequently, strictly periodic signals that persist indefinitely cannot be robustly identified unless the observing duration is sufficiently longer than the period of interest.

By contrast, the detection and characterization of \emph{characteristic variability timescales}—such as correlation times,
transition times, or typical fluctuation scales—represent a statistically distinct problem.
Such timescales do not necessarily correspond to strictly periodic behavior and may be meaningfully inferred even when the observing duration is only a few times longer than the characteristic timescale itself \citep{Priestley1981,KayMarple1981,Shumway2017}.

This distinction is well established in the time-series literature, particularly in model-based approaches such as autoregressive (AR) and state-space models.
In these frameworks, parameter estimation is performed using the likelihood structure of the entire dataset, rather than relying on the assumption of independence among individual samples \citep{Kay1993,Shumway2017}.
In the present study, we explicitly adopt this framework and conduct our variability analysis using state-space modeling and the corresponding AR representation.

Consistent with this theoretical understanding, astronomical time-series analyses commonly restrict the discussion of characteristic timescales or time lags to a fraction of the total observing duration.
For example, \citet{Komossa2022}, in their discrete correlation function analysis of the optical light curve of OJ~287, explicitly restricted reported lag measurements to timescales $\le 1/3$ of the light-curve length.
They further noted that results at timescales between one-third and one-half of the total duration become increasingly unreliable.
Motivated by these theoretical considerations and established astronomical practice, we define the range of variability timescales considered in this work as follows.

For the upper bound, we restrict the analysis to $\tau \le T_{\mathrm{obs}}/3$, where $T_{\mathrm{obs}}$ is the total observing duration of each epoch.
This choice avoids over-interpreting poorly sampled long timescales while retaining sensitivity to physically meaningful variability.
For the lower bound, the light curves are sampled at 10~sec intervals.
Applying standard sampling considerations, we adopt twice the integration time, i.e.\ 20~sec, as the shortest timescale for which variability can be meaningfully discussed.

Accordingly, throughout this paper we confine our analysis and interpretation of variability to the range
\begin{equation}
20~\mathrm{sec} \le \tau \le T_{\mathrm{obs}}/3 .
\end{equation}
Within this defined range, our objective is not to establish strictly periodic signals persisting indefinitely, but rather to characterize statistically supported variability timescales intrinsic to \SGR.

\subsection{Intensity Distribution and Structure Functions}
\label{sect:hist}
\begin{figure}[H]
\includegraphics[trim={4mm 4mm 3mm 4mm}, clip,width=\if@twocolumn\linewidth\else\textwidth\fi]
{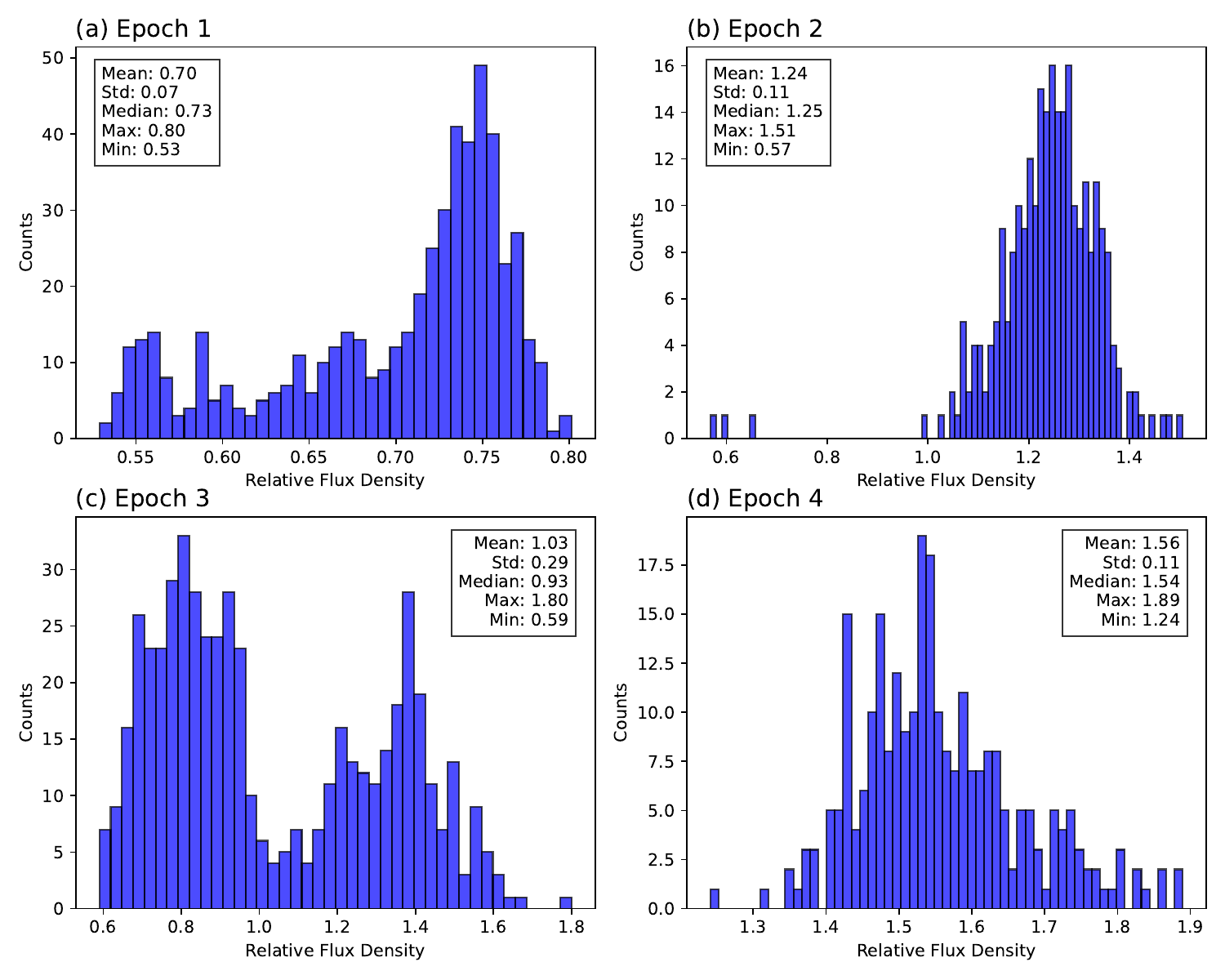}
\caption{
Distributions of the relative flux density of \SGR\ in the four observing epochs (Ep.~1-4).
The bin width in each histogram is set to as one-tenth of the standard deviation of the corresponding dataset.
The legends display key statistical parameters for each epoch.\\
{Alt text: Histograms of the relative flux density for each of the four observing epochs. The distributions are non-Gaussian and differ between epochs, indicating distinct brightness variability patterns in \SGR.}
}
\label{fig:fig06}
\end{figure}

We examine the statistical properties of the relative flux-density variations of \SGR.
Figure~\ref{fig:fig06} presents the distributions of the relative flux density for the four observing epochs.
Statistical tests reject the hypothesis of normality and indicate significant differences among the epochs.
 However, the distributions for Ep.~2 and 4 exhibit concentrations around a single representative value.
\makeatletter
\makeatother

To further investigate these characteristics, we  computed  the structure functions (SFs).
The SF quantifies time-series variability as a function of time delay by measuring the mean squared difference between data points separated by a lag $\tau$, thereby characterizing the correlation properties of variability across different timescales.
SFs are commonly examined by fitting a power-law form,
\begin{equation}
\mathrm{SF}(\tau) \propto \tau^\beta ,
\end{equation}
where $\tau$ is the time lag and $\beta$ is the power-law index.
The classification based on the SF slope $\beta$ follows the conventions of \citet{Simonetti1985} and \citet{Press1978}.
Under the assumption of a stationary and self-similar stochastic process, the fitted slope $\beta$ can be interpreted as an empirical indicator of the scaling behavior of the variability.
For a broad class of stochastic processes characterized by a power-law power spectral density (PSD), $\mathrm{PSD}(f)\propto f^{-\alpha}$, the structure-function slope is approximately related to the PSD index through $\beta \simeq \alpha - 1$ for $1 \lesssim \alpha \lesssim 3$.

In this framework, the value of $\beta$ reflects the degree of temporal correlation rather than uniquely defining a physical noise type:
\begin{enumerate}
 \item $\beta = 0$: white-noise--like behavior (no temporal correlation).
 \item $0 < \beta < 1$: weakly to moderately correlated variability, often described as flicker-like (scale-invariant) behavior.
 \item $\beta \simeq 1$: correlated variability with approximately linear SF growth over the fitted range.
 \item $1 < \beta < 2$: strongly correlated (long-memory) variability, sometimes described as red-noise--like behavior.
 \item $\beta \simeq 2$: random-walk--type behavior (integrated noise), corresponding to a non-stationary limit.
\end{enumerate}

We emphasize that the power-law index $\beta$ provides a phenomenological description of the correlation structure over the range of time lags where the structure function is statistically well constrained, and does not by itself uniquely determine the physical origin of the variability.
SFs are particularly useful for irregularly sampled time series.
In astronomy, structure functions were first applied to flux-variability studies by \citet{Simonetti1985} for extragalactic radio sources.

In the present analysis, we explicitly distinguish between the cases of $\beta \approx 0$ and $\beta > 0$.
Operationally, we refer to variability with $\beta \approx 0$ as white-noise--like behavior and to variability with $\beta > 0$ as temporally correlated (red-noise--like) behavior.

\begin{figure}[H]
\includegraphics[trim={4mm 4mm 3mm 3mm}, clip, width=\if@twocolumn\linewidth\else\textwidth\fi]
{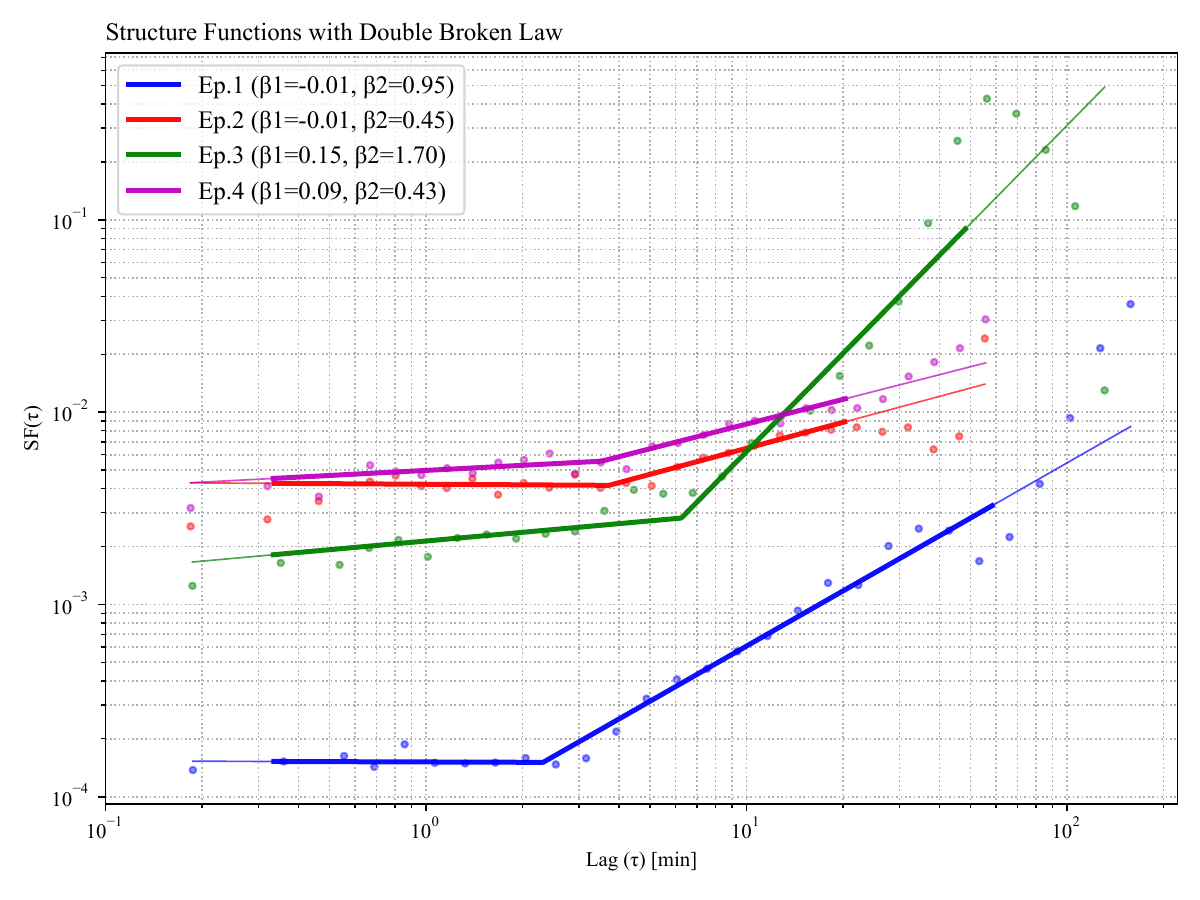}
\caption{
Logarithmically binned structure functions  derived from  the light curves.
The points represent the computed SF values, and the solid lines in the corresponding colors show the best-fit broken power-law models.
The fit is performed over the range $20~\mathrm{sec} < \tau < T_{\mathrm{obs}}/3$, where $T_{\mathrm{obs}}$ denotes the total duration of each observing epoch.
The thick line represents the fitted curve within this interval, and the thin line shows its extrapolation outside this range.
The $\beta$ values shown in the legends indicate the slopes of  the two fitted  power-law segments. 
Note that the absolute  normalization of the vertical axis  is not directly comparable across epochs due to differences in the correlated flux densities of the reference sources. 
Interpretation should therefore focus primarily on the slope of each curve rather than on its absolute amplitude.
\\ {
Alt text: Structure functions of the \SGR\ light curves for all four epochs, plotted on logarithmic scales. 
The curves  are not adequately  described by a single power law and are approximated by a two-segment broken power-law fit. 
White-noise--like behavior is observed on timescales below $2.3$--$6.3~\mathrm{min}$, and the curves transition to red-noise--like behavior at longer lags.
}
}
\label{fig:fig07}
\end{figure}

The structure functions (SFs) derived from our data are shown in Figure~\ref{fig:fig07}.
As discussed above, the interpretation of the SFs is restricted to a limited range of time lags over which the estimates are statistically reliable.

Within this statistically reliable range, the SFs for all epochs are not adequately described by a single power law.
Instead, they exhibit a systematic change in slope with increasing time lag $\tau$, as illustrated in Figure~\ref{fig:fig07}.
Accordingly, we model the SFs using a broken power-law form with two segments, and the best-fit parameters are summarized in Table~\ref{tab:SFs}.\\
~~For each epoch, the SF is approximately flat at short time lags ($\tau \lesssim 2.3$--$6.3~\mathrm{min}$), indicating white-noise--like variability on short timescales.
~~At larger $\tau$ within the fitted range, the SFs exhibit a positive slope, reflecting the emergence of temporally correlated (red-noise--like) variability.
Although the detailed values of the slopes and break times vary among epochs, this two-segment behavior is common to all datasets.
As $\tau$ approaches the upper end of the fitted range, the correlated component becomes more pronounced.
However, our interpretation remains restricted to the statistically validated range shown in Figure~\ref{fig:fig07}.

%
%
\begin{table}[htbp]
\begin{tabular}{lccc}
\hline
Epoch&$\beta_1$&$\beta_2$&$\tau_B$ [min]\\
\hline
Ep.~1&-0.01$\pm$0.66& 0.95$\pm$0.22&2.32$\pm$~2.24\\
Ep.~2&-0.01$\pm$0.30& 0.45$\pm$0.07&3.72$\pm$~2.74\\
Ep.~3&+0.15$\pm$5.73& 1.70$\pm$1.69&6.26$\pm$37.53\\
Ep.~4&+0.09$\pm$0.20& 0.43$\pm$0.17&3.51$\pm$~2.89\\
\hline
\end{tabular}
\captionsetup{width=\columnwidth}
\caption{
Parameters obtained from fits to the logarithmically binned structure functions.
Best-fit power-law slopes ($\beta_1$, $\beta_2$) and break timescales ($\tau_B$) derived from two-segment broken power-law models.
}
\label{tab:SFs}
\end{table}
%
\makeatletter
\makeatother
\subsection{Investigation of  variability timescales }
\label{sect:qpo}
Following previous studies of quasi-periodic oscillations (QPOs) in emission from \SGR\ at infrared and radio wavelengths
\citep{Genzel:03, Yusef-Zadeh:06, Hamaus:09, Miyoshi:11}\footnote{See also the references cited in \cite{Kato:10}.},
we applied two methods—the Lomb--Scargle method (LSM) and the state--space model (SSM) approach—to  our unevenly sampled data to search for characteristic time variability.

\subsubsection{ Lomb--Scargle Method }
\label{sect:LSM}
We first applied the Lomb--Scargle method (LSM) to our data to search for characteristic variability timescales.
The maximum values appearing in the spectra for Epochs~1 to 4 are $13.7$, $8.08$, $15.5$, and $9.4$~min, respectively.
However, all spectra exhibit numerous fluctuations across the entire timescale range, and no dominant peaks are identified.
There is therefore no compelling evidence for a specific periodicity.

The LSM is a spectral method designed to mitigate biases arising from uneven sampling. Nevertheless, a noticeable peak is observed around $8$~min in Epoch~3. This feature is most likely attributable to residual sampling effects.
A similar feature also appears near $8$~min in the LSM results for simulated fixed-value time-series data (see Appendix~\ref{Sec:unevensample}, Figure~\ref{fig:fig11}).

The spectra show substantial variability over a wide range of timescales, with pronounced random fluctuations about a nearly flat level at shorter timescales (below $\sim 5$~min).
This behavior resembles a finite-length realization of white noise, in which an underlying flat spectrum is accompanied by large point-to-point fluctuations.

\begin{figure*}[t]
\centering
\includegraphics[trim={0.0cm 0.00cm 1.25cm 0.0cm},clip,width=\textwidth,height=180mm]{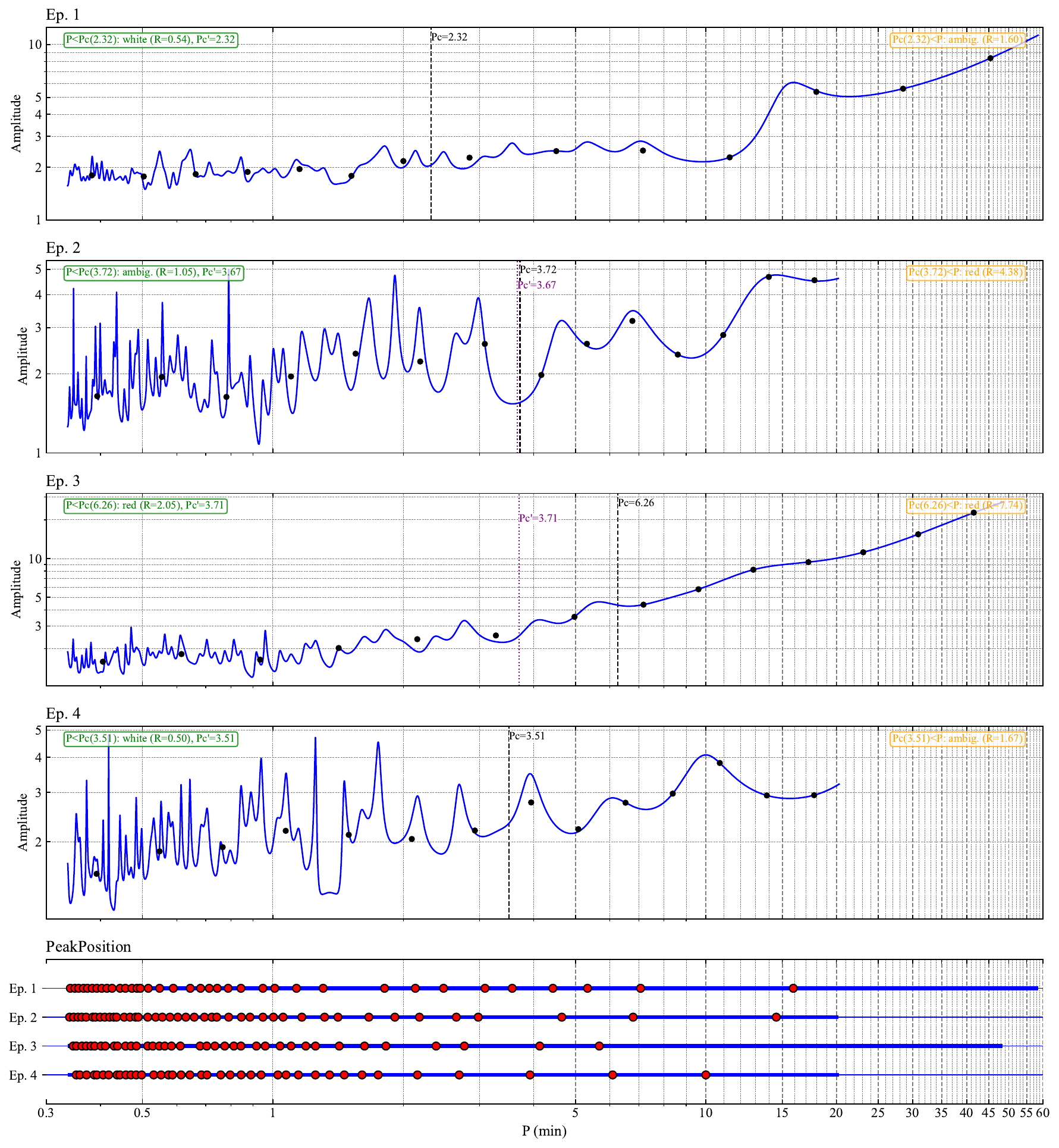}
\caption{Spectra derived from the optimized autoregressive (AR) models based on SSM. The top panels show the spectra for each of the four epochs obtained from the optimized AR models. 
Note that  we omit plots at timescales larger than $ T_{\mathrm{ obs}/3}$ since our observing durations are insufficient to assess spectral features robustly at longer timescales. 
Peaks (local maxima) in the spectra are identified from each spectrum.
In the bottom panel, their positions are plotted as red circles along line segments corresponding to each epoch.
The numerical values and classifications obtained from the DR-WR test, applied separately to the spectral ranges below and above \(P_{\rm c}\), are shown at the upper left and upper right of each panel. The black dots show the median amplitudes in each bin.See also the main text in Section~\ref{sect:psdwhite}.
\\ { Alt text: Spectral power distributions derived from autoregressive analysis based on the SSM. Top: Spectra for the four epochs. Bottom: Locations of spectral maxima positions.}}
\label{fig:fig08}
\end{figure*}
\begin{figure*}[t]
\centering
\includegraphics[trim={0.0cm 0.00cm 1.25cm 
0.0cm},clip,width=\textwidth,height=180mm]
{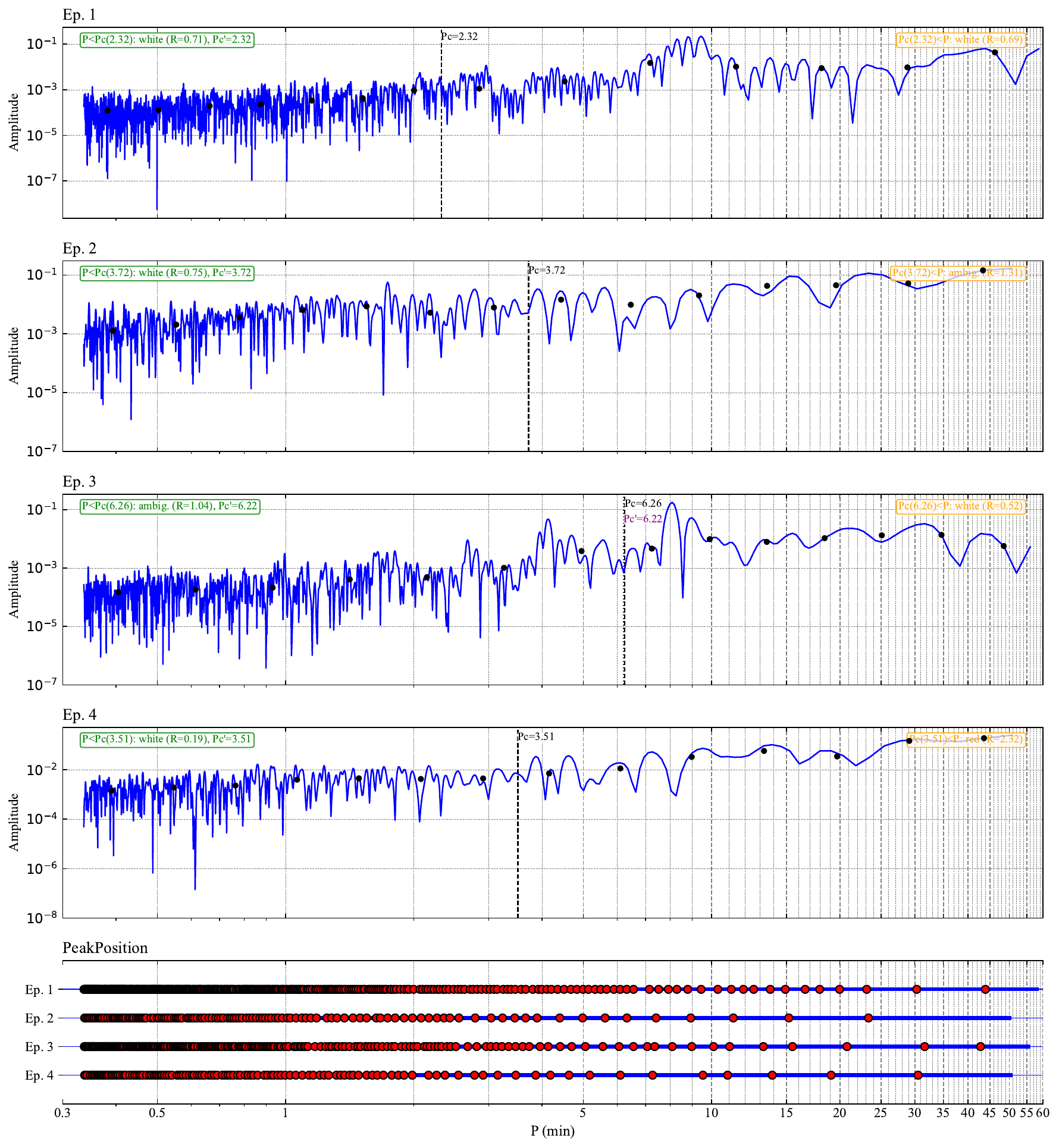}
\caption{
Spectra derived from the Lomb--Scargle method. 
The top four panels show the spectra for each of the four epochs.
Timescales longer than $T_{\mathrm{obs}}/3$ are omitted, since the observing durations are insufficient to robustly assess spectral features at larger timescales.
Peaks (local maxima) in the spectra are identified from each spectrum.
In the bottom panel, their positions are plotted as red circles along line segments corresponding to each epoch.
The numerical values and classifications obtained from the DR-WR test, applied separately to the spectral ranges below and above \(P_{\rm c}\), are shown at the upper left and upper right of each panel. 
The black dots show the median amplitudes in each bin.
See also the main text in Section~\ref{sect:psdwhite}.
\\
{\footnotesize\textit{Alt text:} Spectra derived from the Lomb--Scargle method. 
Top: Power spectra for the four epochs. 
Bottom: Peaks (local maxima) in the spectra are plotted as red circles along line segments corresponding to each epoch.
}
}
\label{fig:figLSM}
\end{figure*}

\subsubsection{ State-Space Model (SSM) }
\label{sect:SSM}
Although the LSM is widely used in astronomy for unevenly sampled time series, it has known limitations in sensitivity and statistical reliability for periodicity detection (see Section~\ref{sect:intro}).
We therefore employ a state-space–model (SSM)–based time-series analysis as a complementary and independent approach.
 The SSM framework was originally developed by \citet{Kalman:1960}, who formulated the state evolution and observation equations within a recursive filtering scheme.

In principle, the SSM-based approach offers several advantages:
\begin{enumerate}
\item It enables robust parameter estimation in the presence of missing data, thereby minimizing their impact.
\item It  explicitly models signal and noise as separate components within the observations , thereby improving interpretability.\footnote{Further methodological advances allow SSMs to (3) dynamically model complex systems and estimate latent (i.e., unobservable) variables; (4) flexibly incorporate prior knowledge, such as physical constraints and external inputs; and (4) handle non-stationary processes by allowing model parameters to evolve over time.}
\end{enumerate}

These capabilities make SSM-based time-series analysis well suited to addressing the limitations inherent in our observational data and  to enhancing the robustness of spectral estimation. 

We fitted an SSM–based autoregressive (AR) model to the data, scanning orders $p=1$–$100$ and selecting the order that  minimizes the prediction residual variance  as the adopted model. 
The AR formulation was chosen for its simplicity, interpretability, and widespread use in time-series analysis, although future work may identify alternative SSM formulations better suited to these data. 
The optimization procedure selected $p=100$ as the optimal order. 
We then computed the spectrum implied by the optimized AR model and used it to estimate  candidate characteristic variability timescales  in the data.

Note that the AR formulation implicitly permits temporally correlated (red-noise–like) variability in the underlying signal, rather than restricting the variability to purely white noise. 
In contrast, within the SSM framework adopted in this work, the observational noise component is modeled as temporally uncorrelated (white) noise.
(We note that, within this modeling framework, source-intrinsic variability that is temporally uncorrelated may not be completely separable from the assumed white observational noise component.)

Figure~\ref{fig:fig08} shows the spectra derived from the optimized AR models based on SSM analysis for each epoch (Ep.~1--4), together with the locations of local maxima (peak positions) in each spectrum.

In all epochs, the spectra do not exhibit a single dominant, sharp peak. 
Instead, they are characterized by a broad-band component in which the spectral power increases toward longer timescales. 
This behavior is more consistent with temporally correlated variability than with strictly periodic signals. 
Compared with the results obtained using LSM, the SSM-based spectra are smoother overall; however, local spectral peaks become more numerous at shorter timescales.

In addition, local peaks are observed at timescales of approximately 15~min in Ep.~1 and Ep.~2. 
Similar local maxima around $\sim$15~min are also seen in the LSM results for Ep.~1, Ep.~2, and Ep.~3. 
This timescale is close to that of the quasi-periodic variability (QPV) reported by \citet{Genzel:03}; however, the peaks detected in the present analysis appear as local enhancements within a broad-band spectrum and do not constitute compelling evidence for strictly periodic variability.

\subsubsection{White-Noise Features in the Spectra Derived from the LSM and SSM Analyses}
\label{sect:psdwhite}

The short-timescale white-noise behavior detected in the structure-function analysis is also reflected in these spectra.
This is suggested not only by the fact that the short-period components resemble the fluctuations expected from white noise in a finite-duration sample, but also by the following independent test.

As an empirical method for distinguishing white-noise--like spectral shapes from red-noise--like spectral shapes, we introduce the \textit{Dispersion-Ratio White/Red Noise Test} (hereafter the DR-WR test).
This method classifies the spectral shape on the basis of the ratio between the scatter among representative values in logarithmic-period bins and the typical scatter within those bins.
Because this procedure was constructed specifically for the present analysis, we treat it not as an established standard test, but as a heuristic method introduced in this work.

We did not adopt a method that judges whiteness solely from the goodness of fit to a flat line.
The main reason is that, although the theoretical spectrum of white noise is flat in expectation, a single periodogram obtained from a finite-length time series intrinsically exhibits large scatter, and therefore even white noise can show substantial apparent fluctuations.
Accordingly, a poor fit to a flat line does not by itself imply non-white behavior.
Conversely, in a spectrum containing a red-noise component, a limited period range can appear approximately flat on average, even when the spectrum is not truly white.
Thus, whiteness cannot be reliably inferred from agreement with a flat line alone.

In addition, in a Lomb--Scargle method, as well as in periodograms in general, the statistical properties of individual spectral points are affected by the finite observation length, uneven sampling, spectral leakage, and local peaks.
Therefore, the assumptions of independence and homoscedasticity that are implicitly required for a simple least-squares fit are not necessarily well satisfied.
For this reason, fitting a flat line may serve as an approximate auxiliary check of whether the spectrum appears roughly flat on average, but it is not necessarily appropriate as a direct test of white-noise behavior itself.

For these reasons, instead of using the goodness of fit to a flat line, we adopted a method that compares the magnitude of systematic variation along the period axis with the magnitude of the local scatter.
Specifically, within each period range, the spectrum was divided into several bins in logarithmic period space, and for each bin we measured a representative amplitude and the scatter within the bin.
To reduce the influence of outliers, the within-bin scatter was evaluated using a robust scale based on the median absolute deviation (MAD).
The conversion of MAD to the standard-deviation scale follows standard practice in robust dispersion estimation.

We then defined the scatter among the representative values of the bins as the inter-bin scatter, and the typical scatter within the bins as the intra-bin scatter.
Their ratio was defined as the dispersion ratio, \(R\).
This dispersion ratio is an indicator of the magnitude of systematic variation along the period axis relative to the magnitude of the local random scatter.
If the spectrum is white-noise--like and flat, differences among the representative values of the bins remain comparable to or smaller than the intra-bin scatter, and the dispersion ratio becomes small.
In contrast, if the spectrum is red-noise--like and varies systematically with period, the inter-bin variation becomes significantly larger than the intra-bin scatter, and the dispersion ratio becomes large.

In the present analysis, the spectral shape in each region was classified as white, ambiguous, or red on the basis of this dispersion ratio \(R\).
Thus, the DR-WR test does not rely on fitting a specific functional form to the spectrum.
Rather, it evaluates whether the systematic variation along the period axis is large compared with the local random scatter, and thereby determines whether the spectral behavior is white-like or red-like.

Furthermore, when the short-period region, \(P < P_{\rm c}\), was not classified as white, the boundary period \(P_{\rm c}\) was gradually reduced, and the largest boundary period \(P_{\rm c}'\) for which the restricted short-period region became white was searched for.
In this way, we additionally examined whether a white-noise--like flat regime exists when the analysis is restricted to periods shorter than the initially adopted boundary period.

For the underlying ideas of robust dispersion estimation and the statistical interpretation of red-noise periodograms, see, for example, \citet{RousseeuwCroux1993}, \citet{Vaughan2005}, and \citet{Vaughan2010}.

The resulting values of \(R\) and the corresponding classifications are shown at the upper left and upper right of the panels in Figures~\ref{fig:figLSM} and \ref{fig:fig08}.
Here, \(P_{\rm c}\) denotes the period corresponding to the break timescale (\(\tau_{\rm B}\)) derived for each dataset and listed in Table~\ref{tab:SFs}.
The quantity \(P_{\rm c}'\) denotes the upper-period boundary of the white-noise regime re-identified directly from the spectrum, and can therefore be regarded as the spectral counterpart of a break timescale.
In most cases, the short-period region up to the break timescale inferred from the structure-function analysis is classified as white noise.
Even in cases where the classification at \(P_{\rm c}\) is ambiguous, the re-searched region below \(P_{\rm c}'\) still shows white-noise behavior.

\subsubsection{Comparison and Interpretation of the LSM and SSM Results}
\label{sect:hikaku}
A direct comparison of the two spectra reveals a notable common feature: 
in both cases, the short-period region is characterized by a largely featureless, scatter-dominated distribution, consistent with expectations for a white-noise spectrum derived from a finite-length time series. 
The smaller scatter seen in the SSM-based spectrum, compared to that obtained with the LSM, is likely attributable to the SSM framework explicitly separating white-noise components as observational errors.

In summary, our analysis of the 340~GHz ALMA observations of \SGR\ yields the following main results:
(1) no dominant variability at a specific, well-defined period is detected; and 
(2) the variability exhibits white-noise--like behavior at short timescales ($\tau \lesssim 2.3$--$6.3~\mathrm{min}$), while a red-noise--like trend becomes apparent at longer timescales.

\FloatBarrier

\section{Discussion}\label{sect:DISC}
In this section, we interpret the observations and their physical implications, focusing on the short-timescale white-noise behavior.
We first examine whether this behavior is intrinsic to \SGR\ and, if so, what physical information it carries.
As no jet has yet been detected in \SGR, we assume that the 340~GHz emission is dominated by the accretion flow.

\subsection{Is the short-timescale white-noise component intrinsic to \SGR\ variability?}
\label{sec:sf_short_discussion}

\subsubsection{Previous Structure-Function Studies}
\label{sec:sf_Previous-Study}

In \citet{Iwata2020}, the epoch-combined SF increases up to $\tau\sim60$~min, while a recalculated SF excluding long-timescale epochs rises to $\sim20$~min and then flattens around $\sim30$~min, interpreted as a characteristic timescale. No specific claim is made for still shorter timescales.

In \citet{MW2021}, an intrinsic SF is constructed by subtracting an estimated measurement-noise term from the observed SF. They report red-noise--like growth over a statistically reliable range of $\sim30$~s to $\sim25$~min, and note reduced robustness at both shortest and longest lags. Their interpretation is therefore restricted to this reliability window.

Thus, neither study provides a detailed discussion of the extremely short-timescale regime.  
Nevertheless, their published SF plots appear to show short-$\tau$ flattening: approximately \(0<\tau<2.3~\mathrm{min}\) in \citet{Iwata2020} and \(0<\tau<0.3~\mathrm{min}\) in \citet{MW2021}. 
For \citet{MW2021}, this should be treated cautiously because the white-noise component is modeled primarily as measurement noise and subtracted when defining their intrinsic SF.

\subsubsection{Measurement-Method Dependence of the Short-Timescale SF Behavior}
\label{sec:sf_ourStudy}
By contrast, our results show white-noise--like behavior over
\(20~\mathrm{s} < \tau \lesssim 2.3\text{--}6.3~\mathrm{min}\),
indicating that the flat (white-noise--like) regime extends to substantially longer timescales than in those previous studies.
One interpretation is that this difference reflects, at least in part, different levels of residual observational errors in the light curves, rather than solely intrinsic source variability.

Observational errors, including atmospheric fluctuations, can also be timescale dependent, much like source variability (e.g., transitions between white- and red-noise components).
For ALMA, these atmospheric effects on millimeter/submillimeter interferometry have been studied in detail \citep{Carilli1999,Holdaway1992,ALMA2015}.
Because atmospheric fluctuations are temporally correlated rather than purely white, they can imprint red-noise--like components on interferometric data.

If calibration does not sufficiently remove this effect, it can contaminate the derived light curves and thus the inferred structure function (SF).
When ALMA sensitivity and angular resolution are fully utilized, observations detect not only \SGR\ and the Galactic-center mini-spiral but also additional nearby sources.
Therefore, if calibrated data recover only \SGR\ and/or the calibrated visibilities remain well fit by a single point-source model, the calibration is likely still incomplete, and non-source variability may remain in the data.

In addition to standard calibration, we applied four further steps:
(1) visibility self-calibration, enabling 10~s snapshot imaging of the Galactic Center;
(2) use of CC2IM maps (constructed only from CLEAN components) to reduce residual-map noise;
(3) relative-flux measurements of \SGR\ against multiple non-variable sources in the same field to suppress common observational systematics; and
(4) simulation-based correction for flux biases caused by time-dependent \UVC\ coverage (i.e., PSF sidelobe variations).
These procedures produced our final \SGR\ light curve.

Compared with previous analyses, this light curve should more closely trace the intrinsic variability of \SGR\ by more effectively suppressing non-source fluctuations.
The corresponding SF shows clear short-timescale flattening, with a white-noise--like regime extending to
$20~\mathrm{s}<\tau\lesssim2.3$--$6.3~\mathrm{min}$,
i.e., to longer timescales than previously reported.
We therefore interpret this short-timescale white-noise--like behavior as likely intrinsic to the intensity variability of \SGR.

Hereafter, we refer to
$20~\mathrm{s}<\tau\lesssim2.3$--$6.3~\mathrm{min}$
as the short-timescale white-noise regime.

\subsection{Interpretation of the short-timescale white-noise regime}
\label{subsec:white_noise}
\citet{Grigorian2024} interpreted the red-to-white transition at intermediate/long timescales as a finite-event-lifetime effect. In their structure-function analysis, red-noise behavior appears at about 2--6~h for $4\times10^6\,M_\odot$, while the SF flattens at  $\gtrsim$5~h, implying statistically independent variability beyond individual events. Thus, red-noise traces temporally coherent single events, whereas white-noise indicates decorrelation between distinct events.

Our analysis probes the opposite temporal regime, revealing a white-to-red transition at very short timescales.
This raises a complementary question: what process sets the onset of temporal correlations on such short timescales?

White-noise--like behavior indicates statistically independent fluctuations with no systematic temporal evolution. In contrast, red-noise--like behavior indicates memory and temporal correlation.
The short-timescale white-noise regime appears both during strong flux changes and during nearly quiescent phases, implying that the fluctuation component remains effectively independent on sufficiently short timescales, regardless of mean-level evolution.

We therefore interpret this white-noise regime as a pre-correlation interval, during which fluctuations have not yet developed coherent temporal structure. 
Its upper bound provides an empirical estimate of the transition timescale above which
temporally correlated variability begins to emerge.
If this timescale reflects information propagation and coupling of local fluctuations, it implies a characteristic spatial scale. For $M = 4 \times 10^6\,M_\odot$, the observed upper limit $\tau \simeq 2.3$--$6.3~\mathrm{min}$ corresponds to a light-crossing scale of $L \simeq 3.5$--$9.6R_{\rm S}$, where $R_{\rm S} = 2GM/c^2$.
Because correlation propagation in an accretion flow is likely slower than $c$, this estimate is an upper bound. Assuming a more realistic propagation speed $v < c$ (e.g., MHD waves, sound waves, or bulk motions), with $v \simeq 0.1c$--$0.3c$, we obtain $L \simeq 0.35$--$2.9R_{\rm S}$, i.e., an innermost-accretion-flow scale.

This interpretation should, however, be kept phenomenological.
Existing theoretical and numerical studies of black-hole accretion flows have not, to our knowledge, explicitly predicted a physically white power spectrum at the shortest timescales, although they do not necessarily exclude such behavior. Instead, reported results have more commonly shown red-noise--like or broken-power-law variability, with characteristic slope changes associated with dynamical activity in the inner flow (e.g.,\citet{Machida+Matsumoto04,Machida+Matsumoto08}). 
In this sense, the observed transition at $\tau \simeq 2.3$--$6.3~\mathrm{min}$ is more naturally interpreted as the onset of temporally coupled variability than as direct evidence that the short-timescale flat component itself is produced by a specific mechanism such as magnetic reconnection. The latter process may still contribute to the correlated variability that develops on longer timescales, but the physical origin of the decorrelated, white-noise--like regime identified here remains uncertain.
\subsection{Implications for High-Resolution Imaging of Sgr~A$^{*}$}
Intrinsic short-timescale variability makes imaging of \SGR\ difficult~\citep{EHTC2022,Miyoshi2024}.
If the upper limit of the short-timescale white-noise regime marks the interval over which structural evolution is still weak, integrations shorter than this limit may capture quasi-static snapshots.
Thus, interferometric arrays with sufficient instantaneous \UVC\ coverage on 
a few minutes or less timescales could directly trace structural variability in \SGR, rather than relying mainly on long-term averaged images.

\section{Conclusion}\label{sect:CONC}

The high sensitivity of ALMA at 340~GHz, combined with carefully implemented self-calibration, enabled us to achieve 
an effective SNR exceeding $10^{8}$ in snapshot CC2IM maps with an integration time of 10.08~s.
This capability allowed us to image the major objects in the Galactic Center with this short integration time, thereby enabling short-cadence monitoring of \SGR.\\
By measuring relative flux densities with respect to non-variable sources within the same FOV, we traced intensity variations of \SGR while largely suppressing common-mode observational errors arising from atmospheric and instrumental fluctuations.
Furthermore, using imaging simulations to quantify the effects of time-dependent \UVC~coverage and associated PSF variations, and applying the corresponding corrections, we isolated the intrinsic intensity variations of \SGR from non-intrinsic variability components at an unprecedented level.\\
We investigated the characteristic variability timescales of \SGR over the range from 20~s to $T_{\mathrm{obs}}/3$ using a combination of structure-function analysis, LSM, and SSM-based time-series analysis.
We identified a short-timescale flat (white-noise--like) regime spanning
$20~\mathrm{s} < \tau \lesssim 2.3$--$6.3~\mathrm{min}$.
At longer timescales, the variability transitions to red-noise behavior.
This short-timescale white-noise regime appears consistently during both phases of pronounced intensity changes and phases with little apparent variation, indicating that, on these timescales, the fluctuation components behave in a statistically independent manner.\\
We interpret the upper boundary of this short-timescale white-noise regime as an empirical transition timescale above which local, short-lived fluctuations in the accretion flow begin to develop temporal coupling and evolve into organized, temporally correlated variability. In this sense, the observed white-to-red transition may mark the onset of memory-bearing variability in the innermost accretion flow. Existing theoretical and numerical studies have not, to our knowledge, explicitly predicted a physically white power spectrum at the shortest timescales, although they do not necessarily exclude such behavior. Reported results have more commonly shown red-noise--like or broken-power-law variability. The physical origin of the decorrelated, white-noise--like regime identified here therefore remains uncertain. Our result should thus be regarded as an observational constraint on models of black-hole accretion variability: successful models must account not only for the correlated red-noise behavior on longer timescales, but also for the emergence of an apparently decorrelated regime below a few minutes.

\begin{ack}
This paper makes use of the following ALMA data: ADS/JAO.ALMA\#2015.1.01080.S. 
ALMA is a partnership of the European Southern Observatory (ESO; representing its member states), the National Science Foundation (NSF; USA), and the National Institutes of Natural Sciences (NINS; Japan), together with the National Research Council Canada (NRC; Canada), the Ministry of Science and Technology (MOST; Taiwan), the Institute of Astronomy and Astrophysics, Academia Sinica (ASIAA; Taiwan), and the Korea Astronomy and Space Science Institute (KASI; Republic of Korea), in cooperation with the Republic of Chile.
The National Radio Astronomy Observatory (NRAO) is a facility of the NSF operated under cooperative agreement by Associated Universities, Inc. (AUI; USA).
The Joint ALMA Observatory (JAO) is operated by ESO, AUI/NRAO, and the National Astronomical Observatory of Japan (NAOJ).
This work was partly supported by a Grant-in-Aid from the Ministry of Education, Culture, Sports, Science, and Technology (MEXT) of Japan (Grant No. 16K05308).
We thank Dr. Naoko Kato for helpful discussions and assistance in understanding time-series analysis.
We also acknowledge the use of OpenAI's ChatGPT-4o for support in developing Python code for data analysis.
The interpretations and conclusions presented in this work are solely those of the authors.
\end{ack}
%
\appendix

\begin{figure}[H]
\centering
\makebox[\columnwidth][l]{%
 \hspace*{-1mm}
 \begin{overpic}[width=0.95\columnwidth,height=6cm]{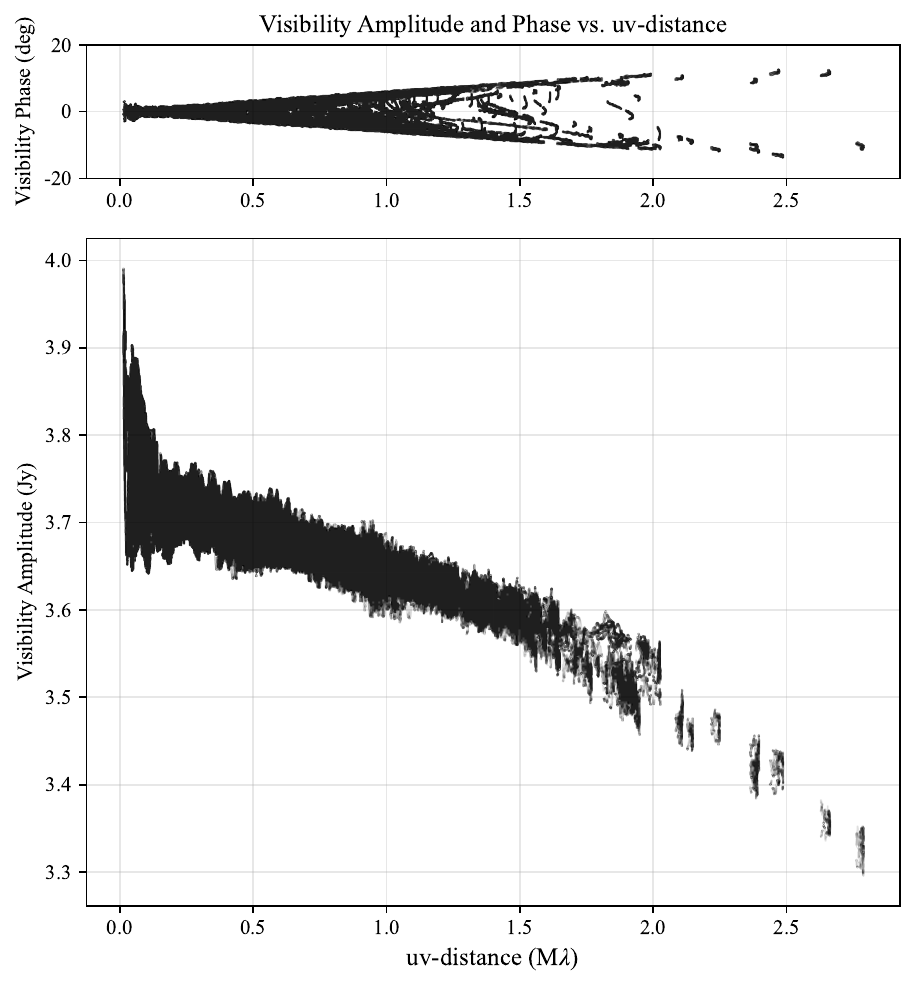}
  \put(-5,90){(a)}
 \end{overpic}
}
%
\begin{overpic}[width=\columnwidth]{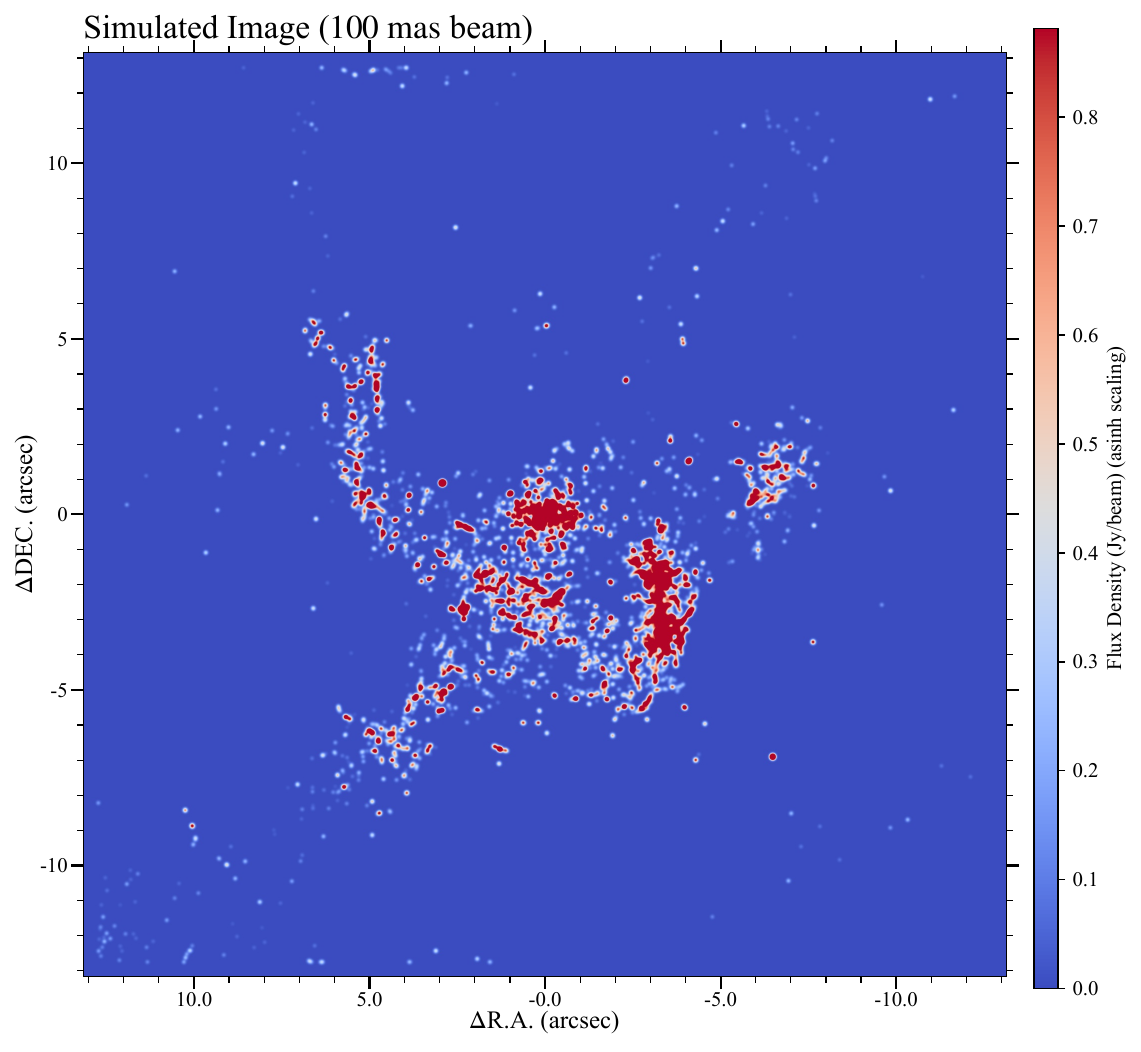}
 \put(-5,90){(b)}
\end{overpic}
\begin{overpic}[width=\columnwidth]{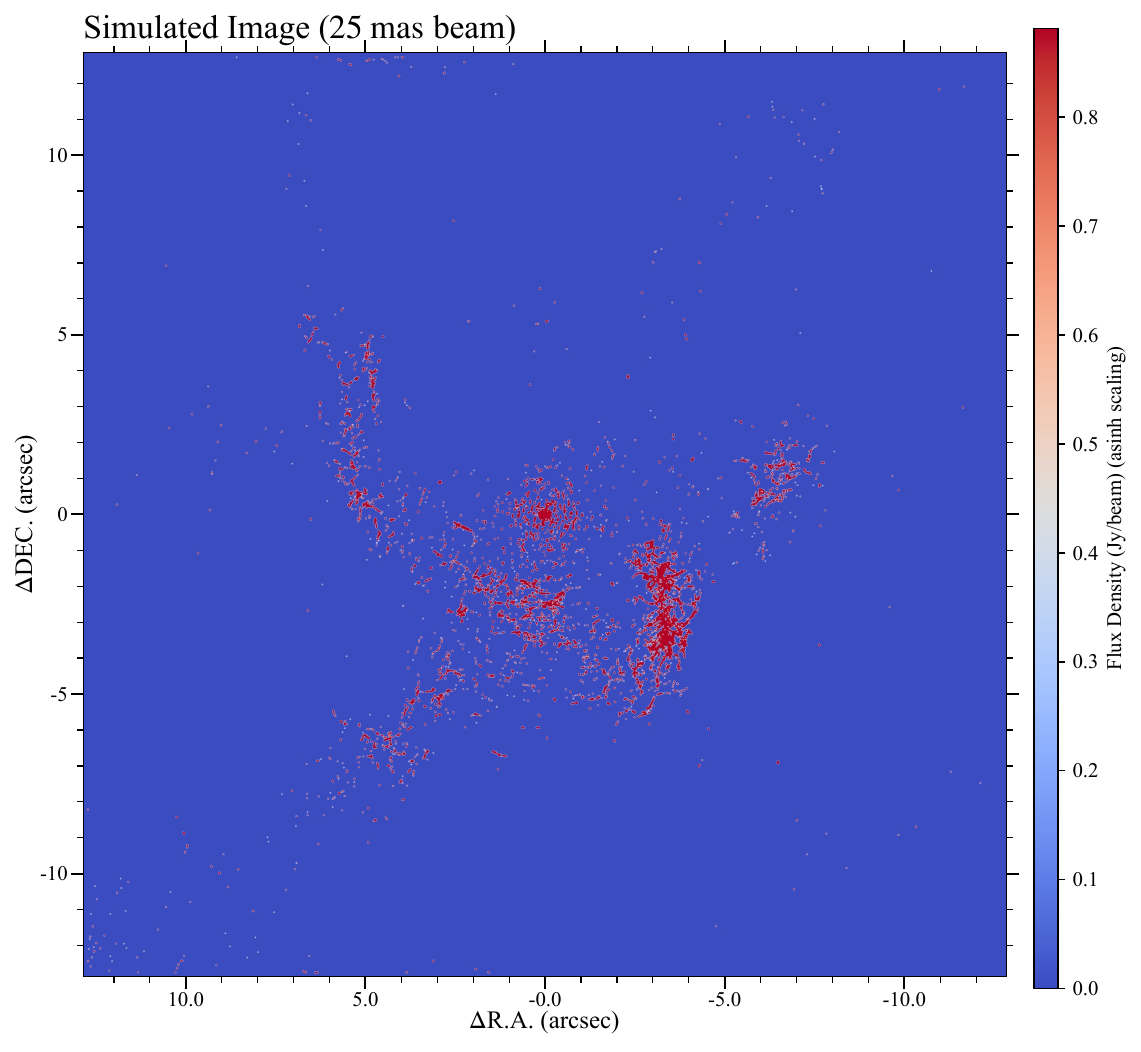}
 \put(-5,90){(c)}
\end{overpic}
\caption{\scriptsize
The visibility-domain quantities and the corresponding simulated images.
(a) Visibility amplitude and phase as a function of \UVC distance.
(b) Simulated image reconstructed with a restoring beam of 100~$\rm mas$.
(c) Same as panel~(b), but with a finer restoring beam of 25~$\rm mas$.
All panels are based on the same underlying source model.
}
\caption*{\footnotesize\textit{Alt text:}
The visibility-domain quantities and the corresponding simulated images.
(a) Visibility amplitude and phase as a function of \UVC distance.
(b) Simulated image reconstructed with a restoring beam of 100~$\rm mas$.
(c) Same as panel~(b), but with a finer restoring beam of 25~$\rm mas$.
All panels are based on the same underlying source model.
}
\label{fig:static-sim-data}
\end{figure}

\section{Imaging Simulation Using a Static Image Model}\label{Sec:CLEAN-sim}
This section presents imaging-simulation study with synthetic data that have no intrinsic time variability. 
The simulation was designed to address three specific objectives:

(1) To evaluate the impact of time-dependent \UVC sampling on snapshot imaging.
In snapshot maps, the \UVC coverage changes with observing time; consequently, 
the dirty beam (point-spread function; PSF) also varies.
Because snapshot \UVC coverage is sparser than that of full-track observations, sidelobes can be relatively prominent with respect to the main lobe, increasing the chance of errors in reconstructing CLEAN components.
As a result, even when the target source is intrinsically static, the images reconstructed from individual snapshots may exhibit spurious variability, biasing flux-density measurements toward an apparent time variability.
 To quantify this effect, we used static model data and applied the same snapshot mapping as well as flux-density measurement procedures as for the 
real data.

(2) To assess the effectiveness of using images composed only of CLEAN components (hereafter CC2IM) versus standard CLEAN maps.
Since the residual map after CLEAN deconvolution typically contains thermal noise, we consider that using images composed only of CLEAN components (CC2IM) can, in principle, provide flux-density measurements with higher signal-to-noise ratios. 
However, in practice, the CLEAN algorithm may identify sidelobe 
related structures as signal and include them among the CLEAN components, thereby introducing noise or artifacts. 
Therefore, whether to use a standard CLEAN map or a CC2IM image can depend on several factors, such as the specific \UVC ~coverage, the structure of the target source, and the instrumental sensitivity. In this study, we examine which approach is preferable under our observational conditions by evaluating imaging-simulation results based on static input data.

(3) To reliably measure the intensity of a variable object, we referenced the sum of the intensities of multiple non-variable celestial objects visible within the same image. Ideally, the reference source would be a point source in addition to being time-invariant. However, because numerous objects are detected in the Galactic Center images, it is difficult to find an isolated point source separable from the others. Therefore, we used the sum of the intensities measured from multiple box regions. 
Using this static image simulation, we compare measurements from a single BOX region near a single object with those from the sum of BOX regions.
As shown in Figure~\ref{fig:fig04} (bottom), the summed compact-reference time series closely tracks that of a single compact source (e.g., IRS~13E) while attaining a higher signal-to-noise ratio; we therefore adopt the summed reference in the denominator of the relative-flux ratio.
%
\subsection{ Simulated Static Data}\label{sec:sim-static-data}
We first describe the static model image used for the simulation. 
To simulate the structure of the Galactic Center as we observed, we constructed a model based on the CLEAN maps obtained from the actual observations.
Specifically, we used the CLEAN maps derived from the full dataset of each epoch, as shown in the second column from the left in Figure~\ref{fig:fig02}. 
From these CLEAN components, we excluded those with negative flux densities, scaled the remaining components by a factor of 1/4, and combined the resultant CLEAN components.
This combined CLEAN component set was then used with the \texttt{UVMOD} task in AIPS to generate simulated visibilities corresponding to the \UVC coverage.

Figure~\ref{fig:static-sim-data} shows the constructed simulated static dataset.
The top panel presents the relationship between projected baseline length and the visibility amplitude and phase.
From this panel, it is evident that fitting either the amplitude or the phase with a flat (constant) line is inappropriate.
If the source were a pure point source, both quantities would be expected to remain constant as a function of projected baseline length.
The lower panels display images reconstructed with different spatial resolutions.
These images clearly reveal the presence of numerous objects in addition to \SGR.
This demonstrates that approximating the Galactic Center data obtained with ALMA as a single point source would lead to substantial discrepancies.
This clearly demonstrates that treating the Galactic Center emission observed with ALMA as a single point source is an oversimplification and would introduce significant systematic errors.

\textbf{Using this set, the \texttt{UVMOD} task produced simulated visibilities matched to the time-dependent \UVC of the real time series; we refer to these as the static simulated visibilities.}

\subsection{Apparent Flux Variability from Simulated Static Data}\label{sec:sim-variation}
\begin{figure*}[tbp]
\centering
\captionsetup{justification=raggedright, singlelinecheck=false}
\includegraphics[trim={0.35cm 0.25cm 0.3cm 0.50cm},clip,width=\linewidth,height=170mm]{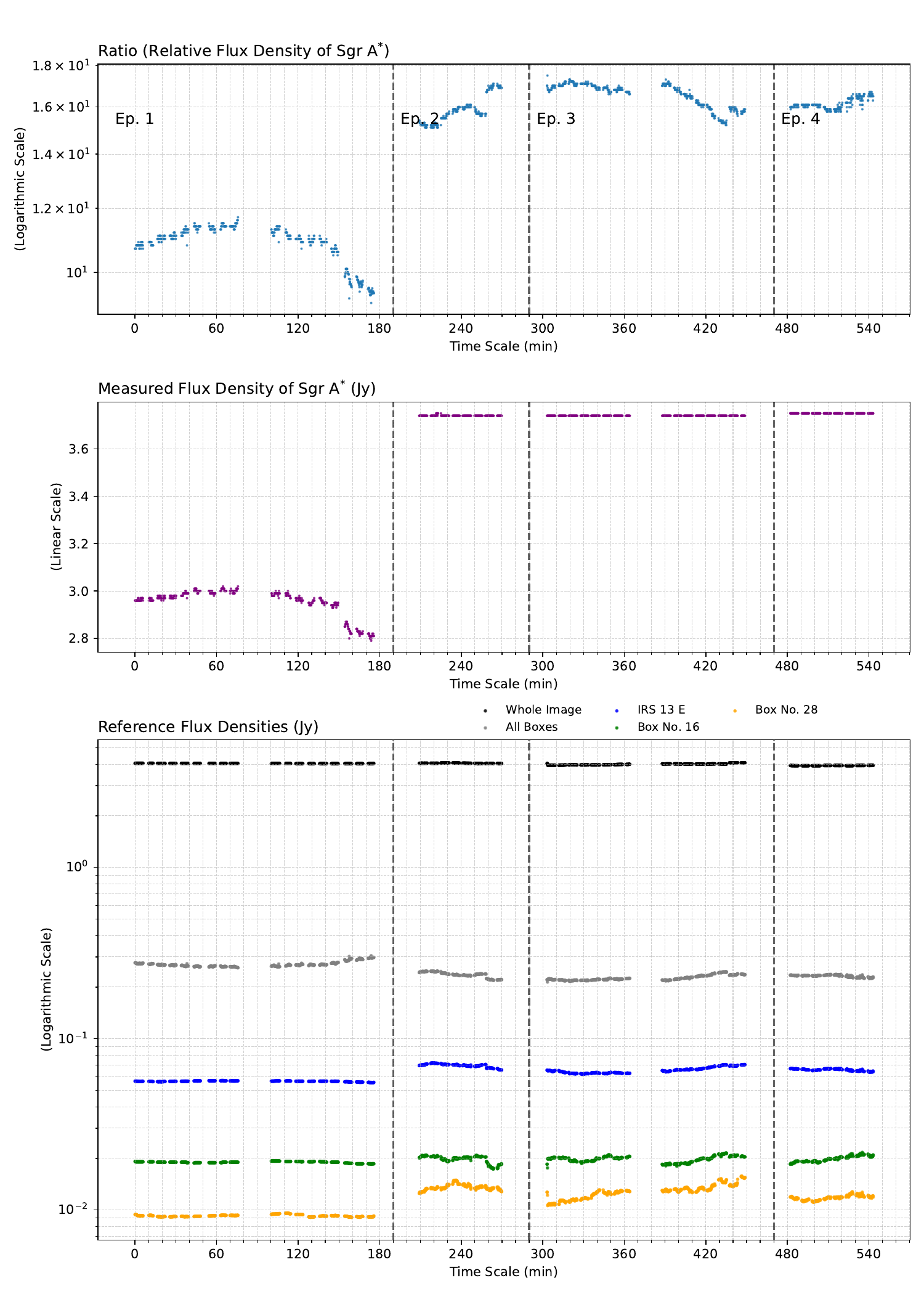}
\caption{
Measurement from snapshot CLEAN maps derived from simulated static visibility data:
The top panel shows the relative flux density of \SGR, which defined as the ratio of the flux density of \SGR~to that of the reference sources. 
The second panel from the top shows the measured flux density of \SGR.
While the bottom panel shows 
(1) the sum of the measured flux densities of the reference box areas, 
(2) those of some box areas (whose locations are illustrated in Figure~\ref{fig:fig01}), and 
(3) the measured flux density of the entire map.
\\ {Alt text: Measured flux densities from snapshot CLEAN maps derived from simulated static visibility data. Time series plots of the measured flux densities for \SGR~and reference regions. Top: relative flux density of \SGR~, showing the variability. Middle: measured flux density of \SGR~, which remains nearly constant. Bottom: Flux densities of selected reference sources and total image flux, showing pronounced variability.}
}
\label{fig:fig09}
\end{figure*}

For this simulated visibility dataset, we performed snapshot imaging using the same procedures as those applied to the real observations, and produced CLEAN and CC2IM maps. flux-density measurements were then carried out using the same method as for the real data. 
The results are shown in Figure~\ref{fig:fig09} for CLEAN images and Figure~\ref{fig:fig10} for CC2IM images. 

As shown, even though the input model is completely static, the flux measurements obtained from both types of snapshot maps exhibit apparent (spurious) variability. 
This variability arises from differences in \UVC coverage and associated changes in the dirty beam (point-spread function; PSF) across the snapshots.
Table~\ref{tab:sim-CLN-CC2IM-measure} shows that the average level of this spurious variability is approximately $3.5\%$ at the $1\sigma$ level, and the amplitude of variability differs across epochs. These differences are likely attributable to differences in the \UVC coverage for each epoch. The largest variability is about $5\%$ in Ep.~1, while the smallest is about $1.4\%$ in Ep.~4.
These results suggest that, when analyzing real data, it is necessary to account for (and, where possible, correct for) the apparent flux variability caused by differences in snapshot \UVC coverage, i.e., variations in the dirty beam (PSF) structure.

\subsection{Comparison of Measurement Results from Two Types of Images}
We now evaluate the extent to which the CC2IM images we adopted 
may be advantageous relative to standard CLEAN maps in the context of this measurement.
According to Table~\ref{tab:sim-CLN-CC2IM-measure}, the average apparent variability defined as "$\sigma$/mean" across the four epochs is approximately $25\%$ smaller when measured from the CC2IM images than from the standard CLEAN maps. 
This indicates lower apparent variability for CC2IM in our dataset.
However, when examining individual epochs, the CLEAN maps yielded smaller variability in Epochs 2 and 3, whereas the CC2IM images resulted in smaller variability in Epochs 1 and 4. 
These results suggest that the optimal image type for flux-density measurement depends on the specific circumstances (e.g., \UVC coverage and source structure).
Based on the simulation results of this study, we found that the CC2IM images provide more stable measurements on average. 
Therefore, we adopted CC2IM images for our flux-density measurements and applied simulation-informed corrections for apparent variability (using the epoch-wise CC2IM-based estimates).
\subsection{
Comparison of Single-Box and Multi-Box Reference Intensity Measurement
}\label{sec:box-ref}
\begin{figure*}[tbp]
\centering
\captionsetup{justification=raggedright, singlelinecheck=false}
\includegraphics[trim={0.35cm 0.25cm 0.3cm 0.50cm},clip,width=\linewidth,height=170mm]{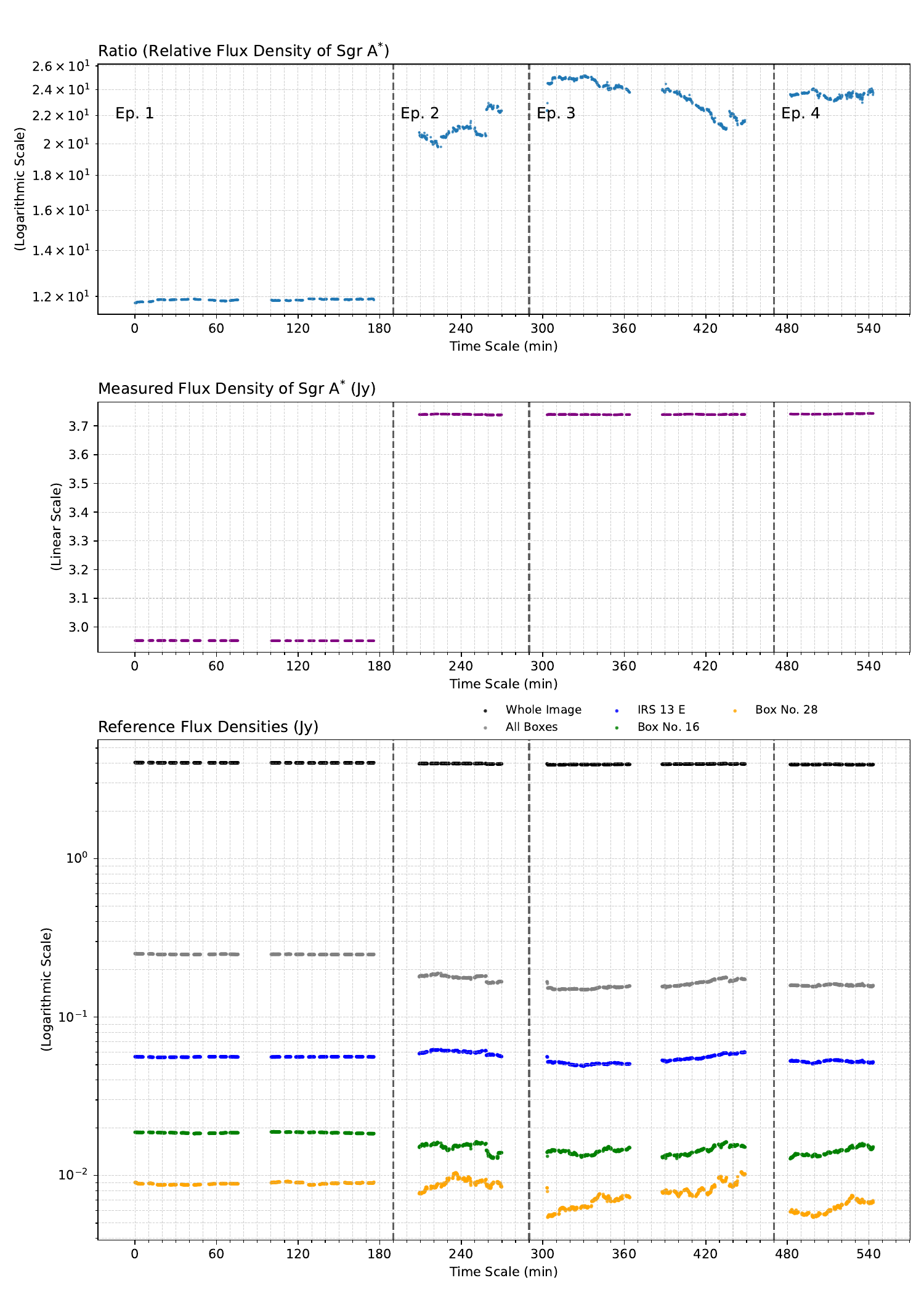}
\caption{ 
Measurement from snapshot CC2IM images derived from simulated static visibility data:
The top panel shows the relative flux density of \SGR, which defined as the ratio of the flux density of \SGR~to that of the reference sources. 
The second panel from the top shows the measured flux density of \SGR.
While the bottom panel shows 
(1) the sum of the measured flux densities of the reference box areas, 
(2) those of some box areas (whose locations are illustrated in Figure~\ref{fig:fig01}), and 
(3) the measured flux density of the entire map.
\\ {Alt text: 
Measured flux densities from snapshot CC2IM images derived from simulated static visibility data.
Time series plots of the measured flux densities for \SGR~and reference regions. Top: relative flux density of \SGR~, showing the variability. Middle: measured flux density of \SGR~, which remains nearly constant. Bottom: Flux densities of selected reference sources and total image flux, showing pronounced variability.}
}
\label{fig:fig10}
\end{figure*}

Table~\ref{tab:refstats} shows the intensity statistics of the measured from the CC2IM images using static simulated data.
Specifically, it presents the statistics for each measurement shown in Figure~\ref{fig:fig10}.
Using this, we compare individual BOX measurements and the sum of all BOX area.
Among the examples of individual BOX measurements, namely IRS~13E, BOX~No.16, and BOX~No.28, the case of IRS~13E exhibits the smallest variation (CV) (average $CV = 2.555 × 10^{-2}$).
On the other hand, the total intensity measured across all BOX regions has a CV of $2.661×10^{-2}$, showing slightly lower stability.
However, when considering intensity through the “CV/Mean” value, IRS13E's BOX has a value of $4.656×10^{-1}$, while the total sum of all BOXes has a value of $1.593×10^{-1}$, approximately three times better.
Considering the significant impact of thermal noise in real data, higher intensity is generally more reliable. Therefore, we consider that the sum of all BOXes, which yields a smaller CV/Mean value, is a better reference source for intensity comparison.
\begin{table*}[tbp]
\centering
\begingroup
\setlength{\tabcolsep}{6pt}    
\renewcommand{\arraystretch}{1.15}
\begin{tabularx}{\textwidth}{lccc|lccc|c}
\toprule
\multicolumn{4}{c|}{\textbf{CLEAN}} &\multicolumn{4}{c|}{\textbf{CC2IM}} & 
\textbf{ratio of variation} \\
\cmidrule(lr){1-4}\cmidrule(lr){5-8}\cmidrule(lr){9-9}
    & mean  & $\sigma$  & $\sigma$/mean &
    & mean  & $\sigma$  & $\sigma$/mean & (CC2IM/CLEAN) \\
\midrule
Ep.~1 & 10.871 & 0.570 & 0.052 & Ep.~1 & 11.859 & 0.035 & 0.003 & 0.055 \\
Ep.~2 & 15.878 & 0.614 & 0.039 & Ep.~2 & 21.075 & 0.838 & 0.040 & 1.028 \\
Ep.~3 & 16.567 & 0.559 & 0.034 & Ep.~3 & 23.599 & 1.248 & 0.053 & 1.570 \\
Ep.~4 & 16.143 & 0.231 & 0.014 & Ep.~4 & 23.585 & 0.235 & 0.010 & 0.699 \\
av.  & 14.865 & 0.493 & 0.035 & av.  & 20.030 & 0.589 & 0.026 & \bf{0.759} \\
\bottomrule
\end{tabularx}
\endgroup
\caption{
Comparison of measured flux density variations in CLEAN and CC2IM images reconstructed from a static model visibility dataset.
For each epoch, we report the mean flux density, standard deviation ($\sigma$), and fractional variation ($\sigma$/mean) for both image types.
The CC2IM images contain only CLEAN components, omitting residual contributions.
The final column shows the ratio of relative variation (CC2IM/CLEAN), highlighting the reduced variability in the CC2IM results for most epochs.
}
\label{tab:sim-CLN-CC2IM-measure}
\end{table*}
\vspace{-3mm}
\vspace{-3mm}

\begin{table*}[t]
\centering
\scriptsize
\setlength{\tabcolsep}{4pt}
\begin{tabular}{lllllll}
\toprule
& \multicolumn{1}{c}{Whole Image}
& \multicolumn{1}{c}{\SGR}
& \multicolumn{1}{c}{All Boxes}
& \multicolumn{1}{c}{IRS 13E}
& \multicolumn{1}{c}{BOX No. 16}
& \multicolumn{1}{c}{BOX No. 28} \\
\midrule
\multicolumn{7}{l}{\textbf{\underline{Epoch~1}}}\\
Mean   & $4.035$       & $2.953$       & $2.490\times10^{-1}$ & $5.602\times10^{-2}$ & $1.860\times10^{-2}$ & $8.879\times10^{-3}$ \\
Std   & $4.785\times10^{-3}$ & $2.547\times10^{-4}$ & $7.457\times10^{-4}$ & $1.541\times10^{-4}$ & $1.285\times10^{-4}$ & $1.172\times10^{-4}$ \\
CV    & $1.186\times10^{-3}$ & $8.627\times10^{-5}$ & $2.995\times10^{-3}$ & $2.751\times10^{-3}$ & $6.908\times10^{-3}$ & $1.320\times10^{-2}$ \\
CV/Mean & $2.939\times10^{-4}$ & $2.922\times10^{-5}$ & $1.203\times10^{-2}$ & $4.911\times10^{-2}$ & $3.714\times10^{-1}$ & $1.486$ \\
\\
\multicolumn{7}{l}{\textbf{\underline{Epoch~2}}}\\
Mean   & $3.984$       & $3.740$       & $1.777\times10^{-1}$ & $5.999\times10^{-2}$ & $1.511\times10^{-2}$ & $8.918\times10^{-3}$ \\
Std   & $9.226\times10^{-3}$ & $1.068\times10^{-3}$ & $6.879\times10^{-3}$ & $1.585\times10^{-3}$ & $9.246\times10^{-4}$ & $6.391\times10^{-4}$ \\
CV    & $2.316\times10^{-3}$ & $2.857\times10^{-4}$ & $3.870\times10^{-2}$ & $2.641\times10^{-2}$ & $6.119\times10^{-2}$ & $7.166\times10^{-2}$ \\
CV/Mean & $5.814\times10^{-4}$ & $7.639\times10^{-5}$ & $2.177\times10^{-1}$ & $4.403\times10^{-1}$ & $4.050$       & $8.035$ \\
\\
\multicolumn{7}{l}{\textbf{\underline{Epoch~3}}}\\
Mean   & $3.950$       & $3.740$       & $1.589\times10^{-1}$ & $5.338\times10^{-2}$ & $1.427\times10^{-2}$ & $7.490\times10^{-3}$ \\
Std   & $1.571\times10^{-2}$ & $5.079\times10^{-4}$ & $8.711\times10^{-3}$ & $3.162\times10^{-3}$ & $7.773\times10^{-4}$ & $1.139\times10^{-3}$ \\
CV    & $3.979\times10^{-3}$ & $1.358\times10^{-4}$ & $5.481\times10^{-2}$ & $5.924\times10^{-2}$ & $5.449\times10^{-2}$ & $1.521\times10^{-1}$ \\
CV/Mean & $1.007\times10^{-3}$ & $3.631\times10^{-5}$ & $3.448\times10^{-1}$ & $1.110$       & $3.819$       & $2.031\times10^{+1}$ \\
\\
\multicolumn{7}{l}{\textbf{\underline{Epoch~4}}}\\
Mean   & $3.938$       & $3.742$       & $1.587\times10^{-1}$ & $5.240\times10^{-2}$ & $1.414\times10^{-2}$ & $6.234\times10^{-3}$ \\
Std   & $4.504\times10^{-3}$ & $9.204\times10^{-4}$ & $1.574\times10^{-3}$ & $7.225\times10^{-4}$ & $8.064\times10^{-4}$ & $5.574\times10^{-4}$ \\
CV    & $1.144\times10^{-3}$ & $2.460\times10^{-4}$ & $9.918\times10^{-3}$ & $1.379\times10^{-2}$ & $5.704\times10^{-2}$ & $8.940\times10^{-2}$ \\
CV/Mean & $2.904\times10^{-4}$ & $6.574\times10^{-5}$ & $6.251\times10^{-2}$ & $2.631\times10^{-1}$ & $4.035$       & $1.434\times10^{+1}$ \\
\\
\multicolumn{7}{l}{\textbf{\underline{Average over epochs}}}\\
Mean   & $3.977$       & $3.544$       & $1.861\times10^{-1}$ & $5.545\times10^{-2}$ & $1.553\times10^{-2}$ & $7.880\times10^{-3}$ \\
Std   & $8.557\times10^{-3}$ & $6.879\times10^{-4}$ & $4.477\times10^{-3}$ & $1.406\times10^{-3}$ & $6.592\times10^{-4}$ & $6.132\times10^{-4}$ \\
CV    & $2.156\times10^{-3}$ & $1.884\times10^{-4}$ & $2.661\times10^{-2}$ & $2.555\times10^{-2}$ & $4.491\times10^{-2}$ & $8.159\times10^{-2}$ \\
CV/Mean & $5.433\times10^{-4}$ & $5.191\times10^{-5}$ & $1.593\times10^{-1}$ & $4.656\times10^{-1}$ & $3.069$       & $1.104\times10^{+1}$ \\
\bottomrule
\end{tabular}
\vspace{0.5ex}
\begin{minipage}{0.95\textwidth}\footnotesize
\caption{
Summary statistics for measurements from snapshot CC2IM images derived from simulated static visibility data. 
Listed for each column are the mean flux density (Jy), standard deviation (Std), coefficient of variation (CV = Std/Mean), and normalized variation (CV/Mean).
}
\label{tab:refstats}
\end{minipage}
\end{table*}

\section{Investigation of the Effects of Irregular Sampling Intervals on Spectral Analysis}\label{Sec:unevensample}
\paragraph*{A.1 Background and Objectives}%
\leavevmode\\

Conventional spectral analysis techniques often produce artifacts when applied to time series data with irregular sampling intervals. These artifacts appear as peaks in the spectrum. The LSM has been proposed as a solution to this issue and is widely used. 

\begin{figure}[tbp]
\includegraphics[trim={4.0cm 1.7cm 3.0cm 1.7cm},clip,width=1.2\linewidth,height=150mm]{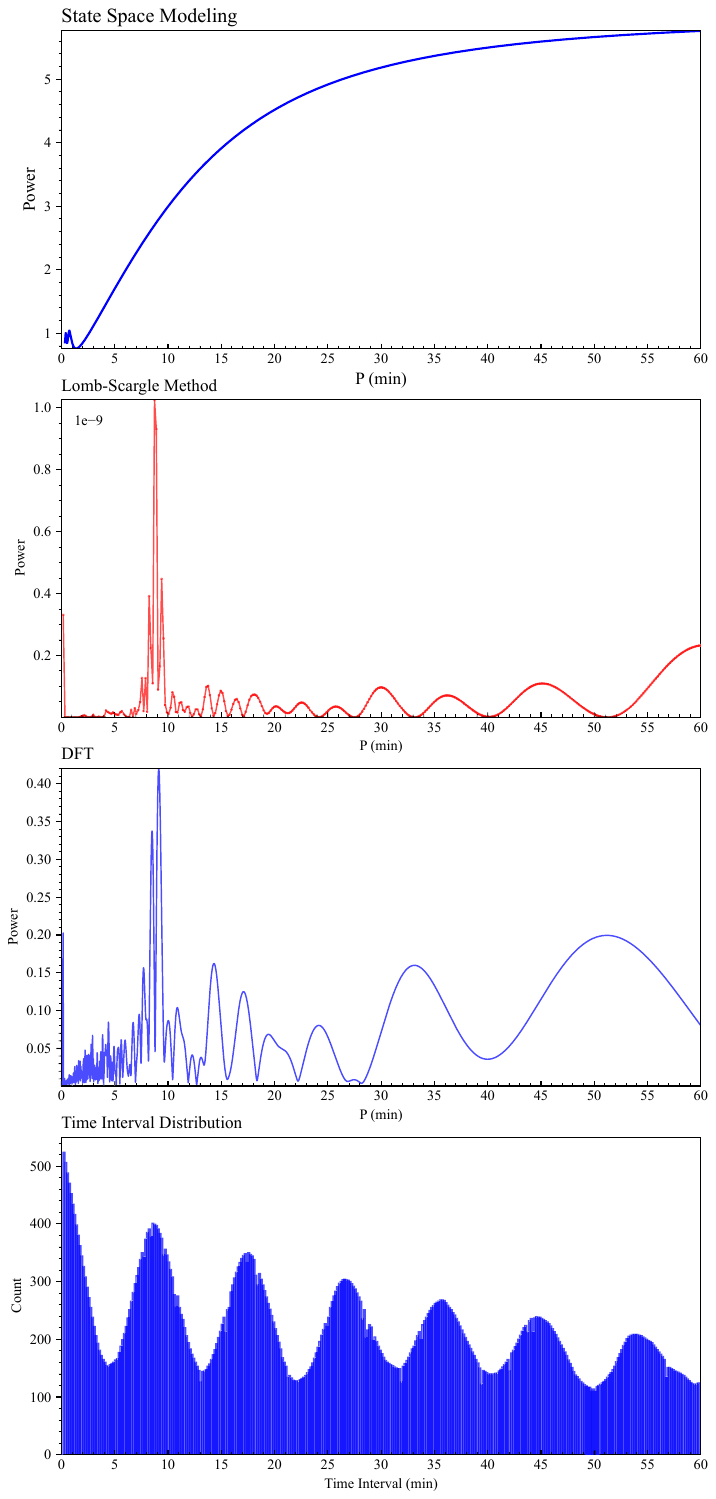}
\captionsetup{justification=justified, singlelinecheck=true}
\caption{%
Spectral analysis results for the Epoch~1 data sampling, evaluated using different methods.
From top to second from bottom, the panels show the results obtained with the state-space model (AR model),
the Lomb--Scargle method, and the discrete Fourier transform (DFT), respectively.
The bottom panel displays the distribution of sampling intervals for the Epoch~1 dataset.%
}
\caption*{\footnotesize\textit{Alt text:} Comparison of spectral analysis methods applied to simulated constant data sampled at the same intervals as Epoch~1. Panels show: state-space model (top), Lomb--Scargle method (middle), and discrete Fourier transform (bottom), along with the histogram of sampling intervals. The SSM method effectively suppresses artifact peaks caused by irregular sampling.}
\label{fig:fig11}
\end{figure}

However, completely eliminating the effects of sampling irregularities (sampling bias) remains challenging.
The ALMA data used in this study were essentially sampled according to a fixed observation sequence; however, due to the inclusion of calibrator observation times, the sampling for the target \SGR~is not perfectly uniform. This appendix explores the implications of such incomplete sampling intervals on spectral analysis.
\paragraph*{A.2 Experimental Design}%
\leavevmode\\
For this experiment, we created a virtual constant time series by assigning the value "1.0" to all data points at the actual sampling times of Epoch~1. 
Given that this series is essentially non-variable, the ideal spectral result would exhibit no peaks.
If peaks appear in the spectrum, it would indicate that they are caused by the sampling distribution itself.
\paragraph*{A.3 Results and Evaluation of Each Method}%
\leavevmode\\
The results are summarized in Figure~\ref{fig:fig11}, which shows the power spectra obtained from three different methods, along with the sampling interval histogram for Epoch~1.
\paragraph*{A.3.1 Discrete Fourier Transform (DFT)}%
\leavevmode\\
Although convenient methods such as interpolating missing data points exist, they were not used here. 
The Fourier transform (DFT) was performed using only the existing data points. 
As a result, clear peaks emerged at the shortest observation interval of 10~seconds and around 9~min, where repeated observations were conducted. 
The sampling interval distribution directly influenced the spectrum.
\paragraph*{A.3.2 Lomb--Scargle Method}%
\leavevmode\\
The LSM is a widely used spectral analysis technique for handling non-uniform sampling. 
However, it has known limitations, which have been discussed previously~\citep{VanderPlas:18}.
It is also unable to fully mitigate the impact of irregular data sampling.
The present results show peaks corresponding to 10~seconds and 9~min, similar to those seen with the direct Fourier transform~(DFT). 
While the power was reduced by several orders of magnitude 
compared to the DFT results, the impact of sampling bias was not fully eliminated, necessitating careful interpretation.
Note that the LSM cannot process a perfectly time-constant series because the variance is zero, rendering the periodogram undefined~\citep{VanderPlas:18}.
Instead, a very slowly increasing series, $y = 1 + \alpha \times t$ with $\alpha = 10^{-9}$, was used for the analysis.

\paragraph*{A.3.3 State-Space Model (AR Model)}%
\leavevmode\\
In the state-space model, an autoregressive (AR) model was applied. 
The order $p$ was varied from 1 to 100, and the model with the smallest residuals, $AR(p=4)$, was found.
The results indicated that no peaks corresponding to the sampling intervals (e.g., 10~seconds) appeared. 
Conversely, minor peaks were detected \aoi around 0.41~min and 0.53~min,  with a steady spectral slope observed across the entire time range. 
These findings suggest that the state-space approach effectively suppresses the generation of spurious peaks caused by the sampling distribution, although some limitations remain.
\paragraph*{A.4 Conclusion}%
\leavevmode\\
This simple test shows that commonly used spectral analysis methods can produce artificial spectral features  based on the way the data is sampled, even if the input signal itself is not variable. 
Among the methods examined, the state-space (AR) model provided the most stable results under the conditions tested, although 
 it is still encounters  certain challenges.
This minimal and illustrative experiment highlights the importance of considering sampling bias when analyzing real observational data.
%

\twocolumn
\kuro

\end{document}